\documentclass[twocolumn]{aastex631}

\usepackage{ amssymb }
\received{May 21, 2024}
\revised{July 18, 2024}
\accepted{July 31, 2024}

\graphicspath{{./}{Plots_final/}}
\begin{document}

\title{A closer look at dwarf galaxies exhibiting MIR variability: AGN confirmation and comparison with non-variable dwarf galaxies}

\correspondingauthor{Archana Aravindan}
\email{aarav005@ucr.edu}

\author[0000-0001-7578-2412]{Archana Aravindan}
\affiliation{Department of Physics and Astronomy, University of California, Riverside, 900 University Ave, Riverside CA 92521, USA}

\author[0000-0003-4693-6157]{Gabriela Canalizo}
\affiliation{Department of Physics and Astronomy, University of California, Riverside, 900 University Ave, Riverside CA 92521, USA}

\author[0000-0002-4902-8077]{Nathan Secrest}
\affiliation{U.S. Naval Observatory, 3450 Massachusetts Avenue NW, Washington, DC 20392-5420, USA}

\author[0000-0003-2277-2354]{Shobita Satyapal}
\affiliation{George Mason University, Department of Physics and Astronomy, MS3F3, 4400 University Drive, Fairfax, VA 22030, USA}

\author[0000-0002-4375-254X]{Thomas Bohn}
\affiliation{Hiroshima Astrophysical Science Center, Hiroshima University, 1-3-1 Kagamiyama, Higashi-Hiroshima, Hiroshima 739-8526, Japan}

\begin{abstract}
Detecting active black holes in dwarf galaxies has proven to be a challenge due to their small size and weak electromagnetic signatures. Mid-infrared variability has emerged as a promising tool that can be used to detect active low-mass black holes in dwarf galaxies. We analyzed 10.4 years of photometry from the ALL$WISE$/NEO$WISE$ multi-epoch catalogs, identifying 25 objects with AGN-like variability. Independent confirmation of AGN activity was found in 68\% of these objects using optical and near-infrared diagnostics. Notably, we discovered a near-infrared coronal line [\ion{S}{9}] $\lambda$ 1.252 $\mu$m in J1205, the galaxy with the lowest stellar mass (log M$_{*}$ = 7.5 M$_{\odot}$) and low metallicity (12 + log(O/H) = 7.46) in our sample. Additionally, we found broad Pa$\alpha$ potentially from the BLR in two targets, and their implied black hole masses are consistent with black hole-stellar mass relations. Comparing non-variable galaxies with similar stellar masses and $WISE$ $W1-W2$ colors, we found no clear trends between variability and large-scale galaxy properties. However, we found that AGN activity likely causes redder $W1-W2$ colors in variable targets, while for the non-variable galaxies, the contribution stems from strong star formation activity. A high incidence of optical broad lines was also observed in variable targets. Our results suggest that mid-infrared variability is an effective method for detecting AGN activity in low-mass galaxies and can help uncover a larger sample of active low-mass ($<$ 10$^{6}$ M$_{\odot}$) black holes in the universe.
\end{abstract}



\section{Introduction} \label{sec:sec1}

The formation of the first black holes (BHs) in the universe is still not well understood \citep{2010MNRAS.409.1022V, 2014GReGr..46.1702N,2020ARA&A..58...27I}. There are two most commonly discussed formation scenarios for black hole seeds, which eventually evolved into supermassive black holes (SMBHs) at the centers of nearly every massive galaxy. In one scenario, the first BHs were formed by the collapse of early Population III stars. The BHs formed by this method are likely to have masses in the range 100 - 1000 M$_{\odot}$ \citep{2001ApJ...551L..27M} and should undergo phases of accretion at super-Eddington rates in order to reach the masses of the SMBH that are currently present in the universe \citep{2005ApJ...633..624V,2016MNRAS.456.2993L}. In an alternate scenario, the first BHs were formed by the direct collapse of gas in the early universe \citep{1998A&A...331L...1S, 2003ApJ...596...34B, 2006MNRAS.370..289B}, and have heavier masses of about 10$^4$ -10$^5$ M$_{\odot}$ \citep{2006MNRAS.371.1813L, 2020ARA&A..58...27I, 2021NatRP...3..732V, 2024NatAs...8..126B}. Other commonly suggested formation mechanisms include runaway collisions in dense star clusters \citep{2004Natur.428..724P, 2012MNRAS.421.1465D, 2021MNRAS.501.1413N}, which can produce BH seeds with masses 10$^3$ - 10$^5$ M$_{\odot}$.

Direct observational evidence for the nature of the earliest BHs is challenging to obtain. However, observing BHs in dwarf galaxies in the nearby universe could help narrow down the formation scenarios \citep{2017IJMPD..2630021M, 2022NatAs...6...26R, 2022MNRAS.514.4912H, 2023MNRAS.523.5610B}. If dwarf galaxies have a higher occupation fraction (percentage of galaxies with BHs) of low-mass ($<$ 10$^3$ M${\odot}$) BHs, it is likely that the BH seeds were formed from the collapse of Population III stars \citep{2010MNRAS.408.1139V, 2012NatCo...3.1304G}. Conversely, if they have a lower occupation fraction of high-mass ($>$ 10$^5$ M${\odot}$) BHs, the direct collapse scenario is favored \citep{2015ApJ...799...98M, 2019ApJ...872..104N, 2020ARA&A..58..257G}.

Detecting active black holes, or active galactic nuclei (AGN), which are actively accreting material, is one of the best methods to determine the presence of BHs in galaxies. Although AGN detections cannot provide a complete census of the BH occupation fraction in dwarf galaxies (since only a small fraction of BHs are likely to be actively accreting and visible as AGN), identifying AGN can still offer constraints on the occupation fraction. Moreover, AGN detections can guide follow-up spectroscopic observations to determine BH masses using methods such as broad-line measurements, further refining our understanding of BH formation scenarios.

With advancements in telescope technology and the advent of large-scale surveys, it has become clear that a sizeable number of dwarf galaxies are likely to host AGN \citep{2013ApJ...775..116R, 2014AJ....148..136M, 2018ApJ...863....1C, 2018MNRAS.478.2576M, 2020ApJ...898L..30M, 2022ApJ...937....7S, 2024MNRAS.528.5252M}. Results from JWST indicate the presence of BH in low-metallicity galaxies in the early universe \citep{2023ApJ...954L...4K, 2023ApJ...957L...7K, 2023A&A...677A.145U, 2023ApJS..265....5H, 2023arXiv230801230M, 2023ApJ...955L..24G} and local dwarfs can be used as laboratories in which to study similar BH, and their effects on the host galaxy in much greater detail than is possible at high $z$ \citep{2018ApJ...863..123B}. Additionally, the detection of AGN in dwarf galaxies can be used to constrain the BH masses and occupation fractions \citep{2023ApJ...946...51C}.

It is challenging to identify active BH in dwarf galaxies using traditional diagnostic methods such as the Baldwin, Phillips \& Terlevich (BPT) diagram \citep{1981PASP...93....5B} or Veilleux-Osterbrock (VO) diagrams \citep{1987ApJS...63..295V} as dwarf galaxies are more likely to have low gas-phase metallicities \citep{1979A&A....80..155L}. \cite{2019ApJ...870L...2C} show that, as the BH mass decreases, the hardening of the spectral energy distribution from the accretion disk changes the ionization structure of the nebula. The enhanced high-energy emission from IMBHs results in a more extended partially ionized zone compared with models for higher-mass BH. This effect produces a net decrease in the predicted [\ion{O}{3}]/H$\beta$ and [\ion{N}{2}]/H$\alpha$ emission line ratios, and the positions of galaxies can shift from the AGN region to the star-forming (SF) region. While diagrams that use [\ion{S}{2}] and [\ion{O}{1}] fluxes are useful for recovering AGN across a range of BH masses and galaxy metallicities  \citep{2019ApJ...870L...2C,2022ApJ...931...44P}, these diagrams still miss a large fraction of AGN \citep{2024arXiv240520312W}, most likely due to the strong contribution from star formation to the line ratios. The low-mass BH in dwarf galaxies also have a small gravitational sphere of influence compared to more massive BH, making them impossible to detect dynamically in galaxies apart from the nearby ones \citep{2014AJ....147...93D, 2015ApJ...809..101D, 2018ApJ...858..118N, 2019ApJ...872..104N}. A low-mass BH, accreting at the Eddington luminosity, will be of comparable luminosity to that produced by stars in the galaxy, also making it challenging to accurately detect AGNs in dwarf galaxies.

A promising method to locate AGNs in dwarf galaxies is by making use of variability. AGNs are known to be variable across the entire electromagnetic spectrum and in different time scales, ranging from months to days depending on the mass of the galaxy \citep{1997ARA&A..35..445U, 2004ApJ...601..692V, 2007AJ....134.2236S, 2010ApJ...721.1014M,2020ApJ...899..136B, 2021Sci...373..789B}. This variability is often attributed to instabilities in the accretion disk or surface temperature fluctuations \citep{2014ApJ...797...19R}. \cite{2020ApJ...896...10B} made use of optical variability using the Sloan Digital Sky Survey \citep[SDSS,][]{2000AJ....120.1579Y} and Palomar Transient Factory \citep[PTF,][]{2009PASP..121.1395L} to identify AGNs in dwarf galaxies. They found that nearly 75\% of low-mass galaxies in their sample with AGN-like variability have narrow emission lines dominated by star formation and would be completely missed by the BPT diagram. Similarly \cite{2020ApJ...889..113M} used the High Cadence Transit Survey  \citep[HiTS,][]{2018AJ....156..186M} to look for optical variability and found that nearly 95\% of variable AGNs in their sample do not exhibit any AGN-like optical emission line ratios. \cite{2022ApJ...936..104W} found 44 out of 25,714 dwarf galaxies had optically variable AGN candidates using optical photometry from the Zwicky Transient Facility \citep[ZTF,][]{2019PASP..131a8002B}. A large sample of optically variable AGN in dwarf galaxies was also obtained from the photometric analysis of various deep field surveys such as Dark Energy Survey \citep[DES,][]{2016MNRAS.460.1270D, 2022MNRAS.516.2736B} and COSMOS \citep{2020ApJ...894...24K}, as well as predictions for possibly uncovering a larger sample from future surveys such as the Rubin Observatory \citep{2019ApJ...873..111I, 2023MNRAS.518.1880B, 2024arXiv240206882B}.

However, high levels of extinction can severely reduce the effectiveness of using optical variability to uncover AGN in low-mass galaxies.  Mid-infrared (MIR) variability, on the other hand, is less sensitive to optically obscured and Compton thick AGN, which form a significant fraction of the low luminosity AGN population \citep{2008ApJ...687..111D, 2015ApJ...815...36A, 2017ApJ...836..165A, 2016ApJ...820....5R}. Ultraviolet and optical emission from the accretion disk is reprocessed by the dust surrounding it and is re-radiated in the MIR. Thus, any fluctuations from the disk would be manifested in the MIR as well, leading to variability in the MIR \citep{2011ApJ...737..105K, 2022ApJ...927..107S, 2023ApJ...950..122L, 2023ApJ...958..135S}.

\cite{2020ApJ...900...56S} found a low incidence of MIR variable dwarf galaxies (0.02\%) using 8.4 years of photometry from the All$WISE$/NEO$WISE$ multi-epoch catalogs and concluded that high cadence data is necessary for MIR selection of AGN in dwarf galaxies. \cite{2022ApJ...936..104W} used forward–modeled MIR photometry of time-resolved $WISE$ coadded images to uncover 148 MIR variable galaxies out of a sample of 79,879 dwarf galaxies. They also found that spectroscopic approaches to AGN identification would have missed 69\% of their $WISE$ MIR variable dwarfs.

While AGNs are variable in the MIR, it is important to realize that there are several other transients that are capable of varying in the infrared as well. Although MIR variability selection is less affected by supernovae (SN) contamination than optical variability selection due to the fainter SEDs of SN \citep{2016PhDT.......177S}, there has been evidence of infrared variability caused by SN and variable stars \citep{2019ApJ...877..110K, 2022MNRAS.513.4057S, 2023ApJ...957...64S}. There have also been instances of infrared transients from interactions between stars and planets \citep{2023Natur.617...55D}, or from mergers between a pair of compact objects such as neutron stars or white dwarfs \citep{2005ApJ...634.1202R, 2017Natur.551...64A}. Although analyzing the evolution of the light curves with time can eliminate non-AGN sources such as rapid, one-off transient events, studies have shown that intermediate-mass black holes are capable of varying over timescales ranging from a few hours to a few days, making it difficult to distinguish them from more rapid transients \citep{2022AJ....163...73S}.

Thus it is important to determine if MIR variability is accurately detecting AGN, specifically in dwarf galaxies. In this work, we aim to determine the effectiveness of MIR variability in tracing AGN activity by selecting a sample of dwarf galaxies that are variable in the MIR from Wide-field Infrared Survey Explorer \citep[$WISE$, ][]{2010AJ....140.1868W} light curves. We then obtained NIR spectroscopic observations of the subset of targets that do not show signs of AGN activity in the optical in order to look for other AGN indicators. These include the presence of NIR coronal lines which have proved to be a useful method to identify the presence of AGN in dwarf galaxies \citep{2021ApJ...911...70B} or broad lines from the broad line region (BLR) that might be too obscured to detect in the optical. We also analyzed how galaxies with variability compare to non-variable galaxies with similar $WISE$ colors to study the differences, if any, between the two samples of galaxies. We do this to understand the effects of any biases in host galaxy properties on the observed MIR variability to indicate if MIR variability is preferentially detected in certain types of galaxies. 

The paper is organized as follows: In Section \ref{sec:sec2}, we describe the variable sample selection, present details of the targets, and describe the observations and data reduction.  In Section \ref{sec:sec3}, we present results of evidence of AGN activity in a subsample of MIR variable dwarfs. In Section \ref{sec:sec4}, we compare the sample of variable dwarfs to a control sample of non-variable dwarf galaxies and analyze the differences between the properties of the two samples. We then summarize our conclusions in Section \ref{sec:sec5}. Throughout the paper, we assume $\Lambda$CDM cosmology, with h = 0.73.

\section{Data and Observations}
\label{sec:sec2}
\subsection{Sample Selection}  \label{sec:sec2.1}

As in \citet{2020ApJ...900...56S}, we created a sample of dwarf galaxies from the NASA-Sloan Atlas (NSA), version \texttt{nsa\_v1\_0\_1}. We started by matching the full NSA catalog to All$WISE$ with $10\arcsec$, which returned a unique match for 637168 out of 641409 (99\%) of objects. We required that the elliptical Petrosian masses and S\'{e}rsic masses differ by no more than 0.5~dex, which \citet{2020ApJ...900...56S} found to effectively remove galaxies cut off in their SDSS images, leaving out 8739 objects.

Because we use catalog All$WISE$ Multiepoch and NEO$WISE$-R PSF photometry for the light curves, we require that the galaxies not be extended or convolved with other sources in $WISE$, respectively, requiring \texttt{ext\_flg<=1}, \texttt{nb==1}. This left 459000 galaxies. Finally, to place our sample on the $WISE$ $W1-W2$, $W2-W3$ color-color diagram, we require that all objects have valid All$WISE$ photometry in the first three bands, with corresponding \texttt{cc\_flags} equal to zero, leaving 316715 galaxies. \citet{2020ApJ...900...56S} found that making a cut of All$WISE$ $W2 < 14.5$~mag removes the Eddington bias from the light curves, leaving 224969 systems. Of these, 6226 galaxies have elliptical Petrosian stellar masses less than $2\times10^{9}$~$h^{-2} M_\sun$, which corresponds to $\log{(M_\star/M_\sun)} < 9.6$ using $h=0.73$. 

\begin{longrotatetable}
\begin{deluxetable*}{ccccccccccc}
\label{table:Table1}
\tablecaption{Summary of AGN confirmations in variable dwarf galaxies}
\tablehead{\colhead{Target} & \colhead{Stellar Mass} & \multicolumn{8}{c}{AGN indicators other than MIR variability} & \colhead{Observed with NIRES}\\
\cline{3-10}
& & Optical BL &  BPT & He II & Optical variability & $WISE$ & NIR Coronal lines & NIR BL & NIR Diagnostics & }
\decimalcolnumbers
\startdata
        J000011.72+052317.4 & 9.29 & \checkmark (a) & $\cdots$  & $\cdots$ & $\cdots$ & \color{blue} \checkmark & $\cdots$ & $\cdots$ & $\cdots$ & $\cdots$\\
        J082912.67+500652.3 & 9.53 &  \checkmark (b,c) &  \checkmark (e)  & $\cdots$ &  \checkmark (g)  & \checkmark (i)  & $\cdots$ & $\cdots$ & $\cdots$ & $\cdots$  \\
        J084702.07+035202.0$^{\bigstar}$ & 9.57 & $\cdots$ & $\cdots$ & $\cdots$  & $\cdots$ &  \checkmark (h) & $\cdots$ & $\cdots$ & \color{blue} \checkmark  & \color{blue} \checkmark  \\
        J085431.18+173730.5 & 9.29 & \checkmark (d) & $\cdots$ & $\cdots$ & $\cdots$ & $\cdots$ & $\cdots$ & $\cdots$ & $\cdots$ & $\cdots$\\
        J090613.76+561015.3 & 9.30 & $\cdots$ & \checkmark (e) & \color{blue} \checkmark  & $\cdots$ & \color{blue} \checkmark & \checkmark (j) & $\cdots$ & $\cdots$  & $\cdots$  \\
        J092810.51+150228.0$^{\bigstar}$ & 9.47 & $\cdots$ & $\cdots$ & $\cdots$ & $\cdots$ & \checkmark (h) &$\cdots$ & \color{blue}\checkmark  & \color{blue}\checkmark  & \color{blue}\checkmark  \\
        J093608.61+061525.9 & 8.29 & $\cdots$ & $\cdots$ & $\cdots$ &$\cdots$ & $\cdots$ &$\cdots$ & $\cdots$  & $\cdots$ & \dag  \\
        J095438.79+403204.4 & 9.52 & \checkmark (a, b) & $\cdots$& $\cdots$ & $\cdots$ & \color{blue} \checkmark & $\cdots$ &$\cdots$ &$\cdots$ & $\cdots$ \\
        J102315.19+122227.9 & 9.01 & $\cdots$ & $\cdots$ & \color{blue} \checkmark & $\cdots$ & $\cdots$ & $\cdots$ & $\cdots$  & $\cdots$ & \dag  \\
        J103206.01+225921.6$^{\bigstar}$ & 8.87 & $\cdots$ & $\cdots$ & $\cdots$ & $\cdots$ & \checkmark (h) & &$\cdots$ & \color{blue} \checkmark & \color{blue} \checkmark  \\
        J114433.55+072643.0 & 9.51 & $\cdots$ & $\cdots$ & $\cdots$& $\cdots$ & \color{blue} \checkmark  & $\cdots$ & \color{blue} \checkmark &  \color{blue} \checkmark  & \color{blue} \checkmark  \\
        J120325.67+330846.1 & 9.10 & $\cdots$ & \checkmark (e) & \color{blue}\checkmark & $\cdots$ & \checkmark (i)  & $\cdots$ & $\cdots$ & $\cdots$ & $\cdots$ \\
        J120503.54+455151.0$^{\bigstar}$ & 7.52 & $\cdots$ & $\cdots$ & $\cdots$  & $\cdots$ & \checkmark (h) & \color{blue}\checkmark & $\cdots$ & $\cdots$ & \color{blue}\checkmark \\
        J122608.78-025226.2 & 8.79 & $\cdots$ & $\cdots$ & $\cdots$ & $\cdots$  & $\cdots$  & $\cdots$  & $\cdots$  &$\cdots$ & \dag \\
        J125305.96-031258.7 & 8.61 & $\cdots$ &$\cdots$ & \checkmark (f) & $\cdots$ & $\cdots$& $\cdots$ & $\cdots$&$\cdots$ & $\cdots$  \\
        J130819.11+434525.6 & 9.57 & $\cdots$& \color{blue}\checkmark & \checkmark (f) & $\cdots$ & \color{blue}\checkmark & $\cdots$ & $\cdots$ & $\cdots$ & $\cdots$  \\
        J132053.67+215510.8$^{\ast}$ & 8.62 & $\cdots$ & \checkmark (e) &$\cdots$ & $\cdots$& \color{blue}\checkmark & $\cdots$ &$\cdots$ &$\cdots$&  \color{blue}\checkmark \\
        J135844.73+123546.9 & 9.34 & $\cdots$ &$\cdots$ & \color{blue}\checkmark & $\cdots$  & $\cdots$ & $\cdots$ & $\cdots$ & $\cdots$ & \color{blue}\checkmark \\
        J140630.09-001939.3 & 8.93 & \checkmark (c) & $\cdots$ & $\cdots$ & $\cdots$ & \color{blue}\checkmark &$\cdots$ & $\cdots$ &$\cdots$ & $\cdots$ \\
        J142835.84+570834.3 & 9.42 & $\cdots$ & $\cdots$ & $\cdots$ & $\cdots$ & $\cdots$ & $\cdots$ & $\cdots$& $\cdots$ & \color{blue}\checkmark  \\
        J143744.58+524333.3 & 8.51 & $\cdots$ & $\cdots$ & \checkmark (f) & $\cdots$ & $\cdots$ & $\cdots$& $\cdots$ & $\cdots$ & \dag  \\
        J152637.36+065941.6 & 9.39 & $\cdots$ & $\cdots$ & \checkmark (f) & $\cdots$ & \color{blue}\checkmark & $\cdots$ & $\cdots$ & $\cdots$ & $\cdots$  \\
        J154404.29+275334.0 & 9.14 & \color{blue}\checkmark & $\cdots$ & $\cdots$ & \checkmark (g) & $\cdots$ & $\cdots$ & $\cdots$ & $\cdots$ & $\cdots$  \\
        J161243.20+124505.2 & 9.26 & $\cdots$ & $\cdots$ &$\cdots$& $\cdots$  &$\cdots$ & $\cdots$& $\cdots$ & $\cdots$ & \color{blue}\checkmark  \\
        J233245.03-005845.8 & 9.45 & \checkmark (a, b) & $\cdots$ & $\cdots$ & $\cdots$ & \color{blue}\checkmark & $\cdots$ & $\cdots$ & $\cdots$ & $\cdots$
\enddata
\tablecomments{\footnotesize Column(1): SDSS name of the target. Column (2): Stellar Mass of the target in units of  $\log{M_\sun}$. Columns(3-10) AGN indicators other than MIR variability. Black checkmarks indicate confirmation in literature (references indicated in parenthesis), while blue checkmarks indicate confirmation based on the analysis carried out in this work based on the SDSS and NIRES spectrum. Column(3): Presence of broad optical Balmer lines  (a)\citep{2006yCat.7248....0V}, (b)\citep{2007ApJ...670...92G}, (c)\citep{2015ApJS..219....1O}, (d)\citep{2016ApJ...820L..19P}. Column(4): Line ratios indicative of AGN activity based on the optical BPT diagram (e)\citep{2013ApJ...775..116R} Column(5): Line ratios indicative of AGN activity based on He II $\lambda$ 4686 \AA diagnostic diagrams (f)\citep{2012MNRAS.421.1043S} Column (6): Presence of optical variability (g)\citep{2020ApJ...896...10B} Column(7): $WISE$ colors indicative of AGN activity (h)\citep{2015MNRAS.454.3722S}, (i)\citep{2017A\string&A...602A..28M} Column(8): Presence of NIR coronal lines (j)\citep{2021ApJ...911...70B} Column(9): Presence of broad NIR Paschen lines Column(10): Line ratios indicative of AGN activity based on NIR diagnostic diagrams. Column(11): Whether the target was observed with NIRES. Only targets that had no other confirmation of AGN in literature (black checkmarks) were observed (see exceptions below)\\
$\bigstar$: Targets that only had $WISE$ colors indicative of AGN activity and no other AGN indicator at the time of proposing for observations.\\
$\ast$ J1320 was classified as an SN candidate in spite of being classified as BPT AGN. \citep{2009ApJ...707.1560I, 2013ApJ...775..116R}.\\
$\dag$ Targets planned for observations but could not be observed due to poor weather. \\Note: These are the reported findings from the literature for the above targets. It is possible that further targeted searches across different wavelengths could reveal additional AGN indicators.}
\end{deluxetable*}
\end{longrotatetable}

\begin{deluxetable*}{ccccccccc}[ht!]
\caption{Observing Log} 
\label{table:Table2}
\tablehead{\colhead{Galaxy} & \colhead{UT Date Obs.} & \colhead{Redshift} & \colhead{Stellar Mass}  & \colhead{Exp. Time\tablenotemark{a}} & \colhead{Slit PA} & \colhead{Airmass} & 
\colhead{Telluric} & \colhead{Flux variation\tablenotemark{b}}\\ [-0.1cm]
\colhead{(SDSS Name)} & \colhead{(YYYY MM DD)} & \colhead{} & \colhead{($\log{M_\sun}$)} & \colhead{(s)} & \colhead{(degrees)} & \colhead{} & \colhead{} &\colhead{(\%)}}
\startdata
        J084702.07+035202.0 & 2023 04 07 & 0.059 & 9.57  & $8\times240$ & 306 & 1.05 & HD 85504 & 15\\ 
        J092810.51+150228.0 & 2023 04 08 & 0.078 & 9.47 & $4\times240$ & 45 & 1.50 & HD 89239 & 25\\
        J103206.01+225921.6 & 2023 04 08 & 0.058 & 8.87  & $4\times240$ & 0 & 1.50 & HD 101060 & 50\\ 
        J114433.55+072643.0 & 2023 04 07 & 0.066 & 9.51 & $4\times240$ & 322 & 1.30 & HD 95126 & 15\\ 
        J120503.54+455151.0 \tablenotemark{c} & 2023 04 07 & 0.065 & 7.52 & $6\times240$ & 0 & 1.17 & HD 95126 & 15\\ 
        J132053.67+215510.8 & 2023 05 08 & 0.022 & 8.62 &  $3\times240$ & 30 & 1.28 & HD 116960 & $<$10\\ 
        J135844.73+123546.9 & 2023 05 08 & 0.024 & 9.34 &  $4\times240$ & 90 & 1.24 & HD 122945 & $<$10\\ 
        J142835.84+570834.3 & 2023 05 08 & 0.059 & 9.42 &  $6\times240$ & 344 & 1.28 & HD 116405 & $<$10\\ 
        J161243.20+124505.2 & 2023 05 08 & 0.034 & 9.26 &  $3\times240$ & 4 & 1.46 & HD 210501 & $<$10
\enddata
\tablenotetext{a}{Exposures were typically done in ABBA nodding.}
\tablenotetext{b}{Variation in flux between individual exposures.}
\tablenotetext{c}{Indicates galaxies with coronal line detections. See Section \ref{sec:sec3.1}}
\end{deluxetable*}

\vspace*{-8mm}

\begin{figure}
\gridline{\fig{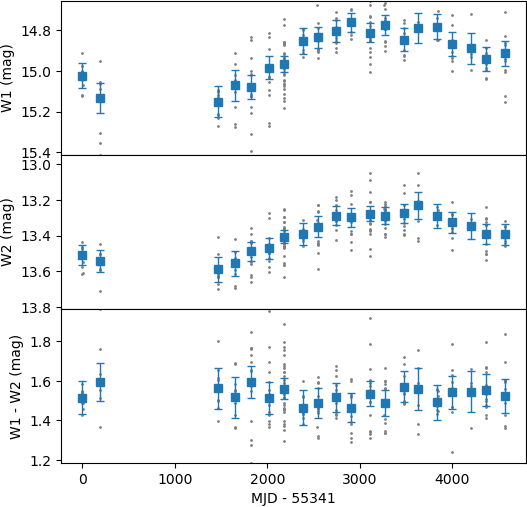}{0.47\textwidth}{}}
\caption{\footnotesize $WISE$ light curves of J1205, one of the variable dwarf galaxies identified in this work, in different $WISE$ bands W1 (3.4 $\mu$m), W2 (4.6 $\mu$m) and the change in color is depicted by W1-W2. The single epoch photometry of the light curve is given by the grey points, while the blue points and the error bars represent the weighted mean and the standard error per 10-day observing period. The subtle redder-when-brighter behavior, consistent with the brightening being due to increasing AGN dominance over the host galaxy, and the gradual nature of the light curves indicate that the variability is likely due to AGN.  \label{fig:fig1}}
\end{figure}

Single-epoch photometry from the All$WISE$ Multiepoch Photometry (MEP) and NEO$WISE$ tables for these 6226 dwarf galaxies were subsequently queried using the All$WISE$ source positions, using zero position offset for MEP and a $3\arcsec$ tolerance for the NEO$WISE$ positions. The single-epoch photometry was likewise cleaned, requiring valid measurements in $W1$ and $W2$, allowing a maximum deviation of $\pm1$~mag between the single-epoch $W2$ and All$WISE$ $W2$ corresponding to the limit set by the Eddington bias, \texttt{na==0} and \texttt{nb==1} for uncomplicated PSF photometry, $W1$ and $W2$ \texttt{cc\_flags} and \texttt{moon\_masked}~$=0$, \texttt{qi\_fact > 0} and \texttt{saa\_sep >= 5}. This removed 11\% of single-epoch measurements, retaining cleaned data for all 6226 dwarf galaxies. On average, there are 276 single-epoch photometric measurements for each source, from which variability metrics were calculated as in \citet{2020ApJ...900...56S}. In that work, we found that Pearson $r$ between $W1$ and $W2$ is the most useful initial discriminant for genuine variability, with $r > 0.4$ being optimal for $WISE$ data. This produced an initial sample of 60 objects, allowing manual inspection of their light curves. We further classified the light curves that exhibit clear variability into those with rapid AGN-like variability and those with supernova-like flares or slow declines in apparent magnitude. 

Although we started with a sample of dwarf galaxies, cutting on objects with AGN-like photometric variability creates a strong selection bias on the otherwise small fraction of contaminating background quasars and galaxies with incorrect spectroscopic redshifts. To mitigate this, we uploaded the $WISE$ coordinates of the variable dwarf galaxy candidates to the SDSS thumbnail server and removed clear instances of nearby background quasars or objects with spectra not matching their catalog redshifts. After removing these contaminants, we were left with a final sample of 25 dwarf galaxies with bona fide, AGN-like, MIR variability (see Fig.~\ref{fig:fig1} for an example). As a check, we estimated the number of background quasars that could be contaminating our sample of 25 objects. Using the SDSS/BOSS quasar catalog \citep{2020ApJS..250....8L}, 98\% of which have $WISE$ photometry (738088 objects), we find that 114119 have $W2 < 14.5$~mag like our sample. As this catalog covers $\sim28\%$ of the sky, the corresponding sky density of SDSS/BOSS quasars is $\sim10$~deg$^{-2}$. However, because our AGN selection criteria is mid-IR based, we correct for the obscured population that may have been largely missed by the SDSS/BOSS surveys. We conservatively assume that about half of the mid-IR AGN-selected sources are heavily obscured \citep[e.g.,][]{2023ApJ...946...27P}, giving a sky density of $20$~deg$^{-2}$. The final sample of 25 objects has a maximum NSA/All$WISE$ position offset of $4\farcs4$, of which 6023 of the 6226 dwarf galaxies satisfy. Then, the expected number of contaminant background QSOs is $\sim0.6$. Thus, even if mid-IR variability selects all background QSOs, it is likely that there is at most one contaminating background QSO in our final sample of 25 dwarf galaxies. The final sample is presented in Table \ref{table:Table1}. 

\begin{figure*}[t!]
\gridline{\fig{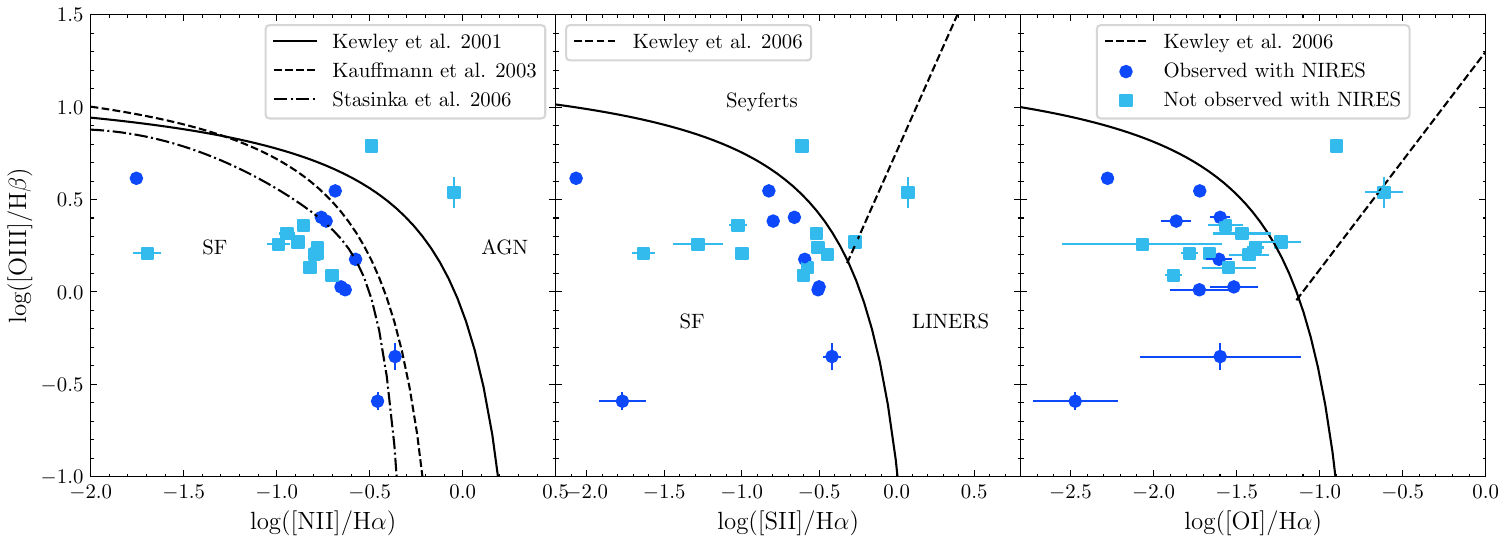}{1\textwidth}{}}
\caption{\footnotesize Optical BPT and VO diagrams showing the positions of the variable targets observed with NIRES (blue) and the variable galaxies not observed with NIRES (cyan). The emission line ratios were derived from fluxes obtained from the SDSS MPA-JHU catalog. Most of the observed variable targets lie primarily in the SF regions, and there is no indicator of AGN activity based on these diagnostic diagrams. \label{fig:fig2}}
\end{figure*}

The stellar masses listed in Table \ref{table:Table1} were obtained from the NSA catalog \citep{2007AJ....133..734B}.  In order to estimate the uncertainty in these masses, we compared the masses listed by NSA with masses reported in other catalogs. We found an average deviation of 0.01 dex between the NSA masses and those reported in the MPA-JHU catalog \citep{2003MNRAS.341...33K, 2007ApJS..173..267S}. This falls in the range of the expected deviations (-0.2 dex - 0.2 dex) between the MPA-JHU and NSA stellar masses determined by \citet{2019ApJ...883...83P}. We also determined an average deviation of 0.3 dex between the NSA masses and those reported by the Portsmouth group \citep{2006ApJ...652...85M} for galaxies with masses determined in all three catalogs. These values provide a measure of the average uncertainty in the masses of our sample. Thus, the stellar masses that we use fall within the range of expected masses of dwarf galaxies even with the predicted uncertainty. Also see Fig. \ref{fig:fig2} for their positions on the optical BPT diagram.

We then extensively searched to see if any variable dwarf galaxies had already been identified as AGN in the literature. Twelve of the 25 (48\%) dwarf galaxies were already identified as AGN by a combination of other optical methods, such as the presence of broad Balmer lines, strong \ion{He}{2} emission, optical variability, and emission line ratios indicative of AGN from the BPT diagram (see Table \ref{table:Table1} for references).

Although the MIR variable galaxy J1320 has BPT line ratios indicative of AGN, it was also classified as a potential SN candidate in the literature \citep{2009ApJ...707.1560I,2013ApJ...775..116R}, so we do not include it among the targets for which AGN activity was previously confirmed. Additionally, only one of the 25 (J1526) variable galaxies was classified as AGN from X-ray observations \citep{2017ApJ...836...20B}. Apart from the references mentioned in the caption of Table \ref{table:Table1}, we also cross-referenced the variable targets in this work with \citet{2014AJ....148..136M, 2022ApJ...931...44P} for BPT AGN, \citet{2018ApJ...863....1C} for optical broad lines, and \citet{2020ApJ...900..124B, 2024AJ....167...31O} for X-ray data, but we did not find any matches.

\subsection{Observations and Reductions} \label{sec:2.2}

We selected a subsample of variable dwarf galaxies for observations in the NIR. The 13 galaxies in the subsample include nine galaxies with no confirmations of AGN by any method (including J1320) and four galaxies that only had $WISE$ colors indicative of AGN activity. These were the only galaxies in the sample exhibiting MIR colors indicative of AGN without any other indicators across the criteria examined. We included the latter four galaxies because young starburst galaxies can often mimic the mid-infrared colors of AGN galaxies \citep{2016ApJ...832..119H} and may not truly be AGNs. Probing the NIR spectra of the variable targets could uncover evidence of AGN activity, such as broad Paschen lines from the BLR or coronal lines indicative of a hard radiation field. 

NIR spectroscopy was obtained on three separate dates (UT 2023 April 7, UT 2023 April 8 and UT 2023 May 8) with the Keck II Near-Infrared Echellette Spectrometer \citep[NIRES;][]{2004SPIE.5492.1295W}. NIRES is a NIR echellette spectrograph with a wavelength coverage across five bands (Z, Y, J, H, K) from 0.94 to 2.45 $\mu$m. There is a small gap in coverage between 1.85 to 1.88 $\mu$m, but this region is dominated by water absorption from the atmosphere and thus has low atmospheric transmission. The average spectral resolution of the five orders ranges between 84 and 89 km s$^{-1}$ (R = 3400). Observations on UT 2023 April 7 and UT 2023 April 8 were taken under variable and heavy cloud cover, while observations on UT 2023 May 8  were taken through light cirrus. Due to unfavorable weather conditions, we could get data for only nine of the 13 targets. Individual exposures for each target were taken for 4 minutes using the standard ABBA nodding, amounting to total exposure times ranging from 12 to 32 minutes, depending on the brightness of the target. A0 spectral class stars were used as telluric standard stars, with measured magnitudes in J, H, and K bands. Observations of the telluric stars were taken before or after each science target to correct for the atmospheric absorption features. A summary of the observations is shown in Table \ref{table:Table2}. 

\begin{figure*}[ht!]
\gridline{\fig{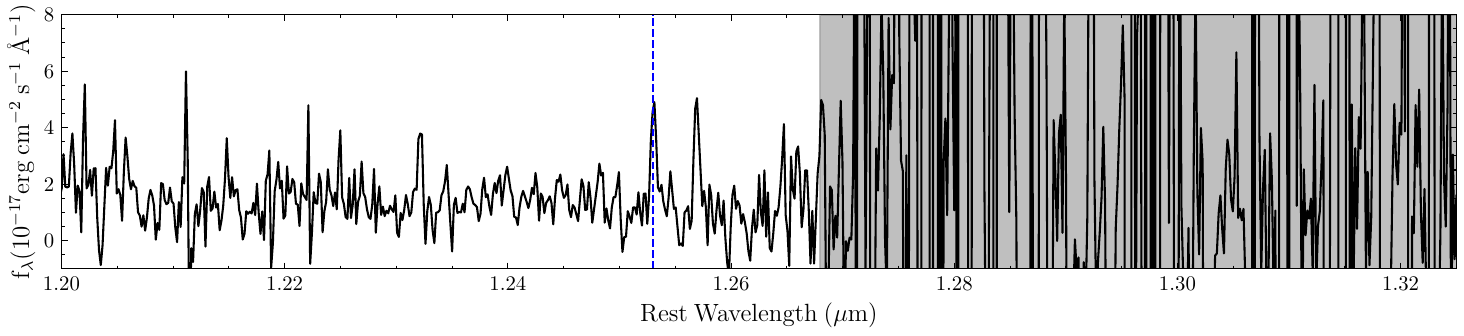}{1\textwidth}{a}}
\gridline{\fig{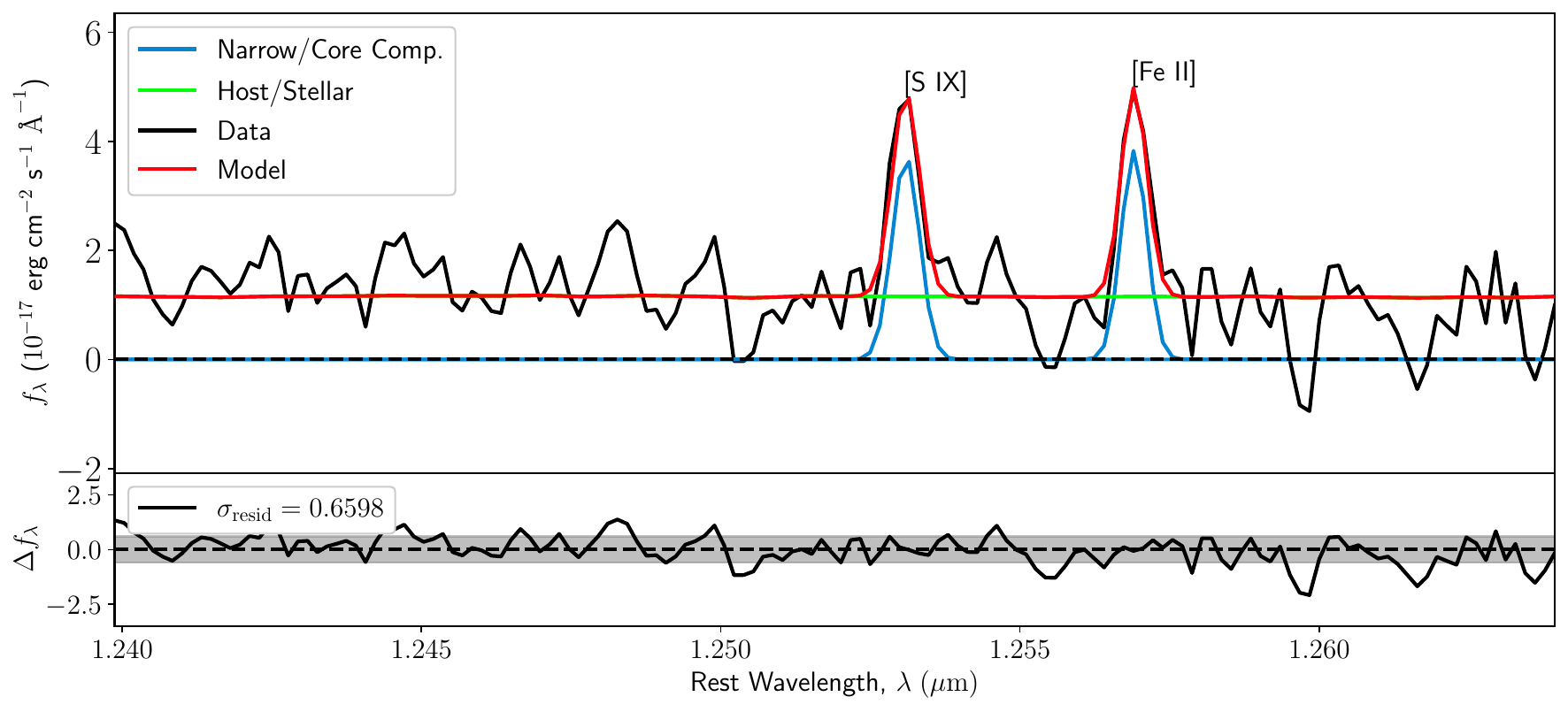}{1\textwidth}{b}}
\caption{\footnotesize (a) NIR spectrum of J1205 in order 5 of NIRES, shifted to rest-frame wavelength. The blue-dashed line indicates the coronal line [\ion{S}{9}] detected at a 5$\sigma$ level. The grey-shaded area represents the region of significant telluric absorption. (b) The zoomed-in spectrum with spectral fits to the [\ion{S}{9}] (1.252 $\mu$m) line and the [\ion{Fe}{2}] (1.257 $\mu$m)line. The lower panel depicts the residuals to the fit, with the 1$\sigma$ error in the spectrum shaded in grey around the dashed line indicating 0. The $\sigma_{\mathrm{residuals}}$ indicates the standard deviation of the residuals determined by subtracting the fit to the spectrum from the data. \label{fig:fig3}}
\end{figure*}

The data were reduced using two modified pipelines. The flat fielding and background subtraction were done by the first pipeline using techniques described in \citet{2003PASP..115..688K} and \citet{2009ApJ...698.1010B}. Rectification, telluric correction, wavelength calibration, and extraction were done with a slightly modified version of REDSPEC \footnote{\url{https://www2.keck.hawaii.edu/inst/nirspec/redspec.html}}. The telluric correction was done by dividing each science target spectrum by the spectrum of its corresponding telluric star and multiplying it by a blackbody curve of the same temperature. Wavelength calibration was done using strong OH lines, and the 1D spectra were then median combined. Flux calibration of individual targets was done by converting the magnitude of the telluric star to the associated flux in that band \footnote{\url{https://irsa.ipac.caltech.edu/data/SPITZER/docs/dataanalysistools/tools/pet/magtojy/}}. A minor corrective factor ($<$5\%) was introduced due to the differences between the center of the NIR bands and the center of the wavelength coverage. Because we observed through clouds, we found 15 - 50 \% variations in the flux levels between individual exposures for the first two nights and less than 10 \% variation in flux on the final night (Flux variations for individual objects are listed in Table \ref{table:Table2}). This introduced uncertainty in the flux calibration, as the latter was done relative to telluric standards observed immediately before or after each science target.  We estimated that the uncertainties are of the order of the flux variations, and we included these added uncertainties in the measurements of fluxes.

\subsection{Spectral fitting}

All NIR spectra were fit using the Bayesian AGN Decomposition Analysis for SDSS Spectra 
\citep[BADASS, ][]{2021MNRAS.500.2871S} code, which was modified to cover the NIR wavelength regime. A narrow component for each emission line was fit simultaneously with a second-order polynomial for the continuum. Absorption features were fit using the penalized pixel fitting method (pPXF, \citet{2004PASP..116..138C}) with templates from the eMILES \citep{2016MNRAS.463.3409V} stellar library. 

The error (noise) in the fit was determined by calculating the standard deviation of the continuum flux, represented by the grey-shaded region in the lower panel of Figure \ref{fig:fig3}
(b). The $\sigma_{\mathrm{residuals}}$, also indicated in the residual plot (lower panel) of Figure \ref{fig:fig3}(b) denotes the standard deviation of the residuals obtained by subtracting the modeled spectrum from the observed data.

\section{Evidence of AGN activity}
\label{sec:sec3}
\subsection{Detection of coronal lines in J1205} \label{sec:sec3.1}
Coronal lines (CL) are high ionization fine-structure lines that arise from collisionally excited forbidden fine-structure transitions in highly ionized species, with ionization potentials that extend well beyond the Lyman limit to several hundred electron volts (eV). They are typically attributed to AGN activity and have been successfully detected in dwarf galaxies hosting AGN \citep[e.g.,][]{2021ApJ...911...70B}
We looked for NIR coronal lines in the nine galaxies observed with NIRES.

The NIR coronal line [\ion{S}{9}] $\lambda$1.252 $\mu$m (I.P. 328.2 eV and critical density log (n$_e$)=9.4 cm$^{-3}$) was detected in J1205. The fit to the spectrum showing the [\ion{S}{9}] is shown in Fig.~\ref{fig:fig3}. The [\ion{S}{9}] line is a 5$\sigma$ detection, and the measured flux of the line is 2.11 $\pm$ 0.52 $\times$ 10$^{-16}$ erg cm$^{-2}$ s$^{-1}$. The full width at half maximum (FWHM) of the line is 138 $\pm$ 10 km s$^{-1}$, which is greater than the spectral resolution of NIRES in that order (85 km s$^{-1}$) and thus is unlikely to be a noise spike or a cosmic ray signal. The errors in the flux and FWHM are given by the standard deviation of the posterior distribution from the MCMC fitting of the lines. Based on the commonly observed NIR emission lines in galaxies \citep{2006A&A...457...61R,2008ApJS..174..282L,2013MNRAS.431.1823M}, no other emission lines are likely to be found at that wavelength.

The detection of [\ion{S}{9}] in J1205 is tantalizing evidence that it likely hosts an AGN. This is the first instance of a NIR coronal line being discovered in J1205, providing strong evidence that J1205 has an active black hole. J1205 has the lowest stellar mass and metallicity among all the targets \citep[log(M$_{*}$) = 7.5 M$_{\odot}$ and 12 + log(O/H) = 7.46;][]{2021MNRAS.508.2556I}. It was classified as a compact starforming galaxy with extremely high [\ion{O}{3}]/[\ion{O}{2}] flux ratios by \citet{2017MNRAS.471..548I} and shown to have a [\ion{Ne}{5}] $\lambda$3426 \AA \hspace{0.09cm} detection and evidence of hard ionization by \citet{2021MNRAS.508.2556I}. Spectroscopic fits to the H$\alpha$ emission line in spectra obtained from the Large Binocular Telescope (LBT) indicate the presence of a broad line feature, which is attributed to expansion motions of supernova remnants \citep{2017MNRAS.471..548I}.  \citet{2023ApJ...945..157H} classified J1205 as a green pea galaxy possibly hosting an AGN based on MIR colors and variability from $WISE$. The low metallicity and stellar mass of J1205 would have made it impossible to detect the presence of the AGN by traditional methods such as the BPT diagram. This indicates that MIR variability can be a promising tool and, sometimes, the only method to detect AGN in low-mass galaxies.

\begin{figure}[h!] 
\gridline{\fig{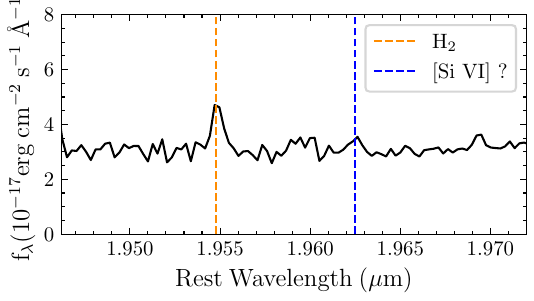}{0.5\textwidth}{}}
\vspace{-5mm}
\caption{\footnotesize 
Keck/NIRES spectrum of J1205 in the wavelength region around [\ion{Si}{6}].  The small emission feature found at the wavelength of [\ion{Si}{6}] (indicated by the dashed blue line) has a flux compatible with the expected flux for [\ion{Si}{6}], however, it is below the 3$\sigma$ detection limit and may not be real.}
\label{fig:fig14}
\end{figure}

While the presence of [\ion{S}{9}] in J1205 is likely to indicate AGN activity, it could also possibly be the result of a tidal disruption event \citep[TDE;][]{ 2023MNRAS.519.2035H,2024MNRAS.528.7076C}. This can be determined by follow-up observations to check if the coronal line persists or declines after an interval of time. However, a TDE would still indicate the presence of a black hole in J1205.

\begin{figure}[h!] 
\gridline{\fig{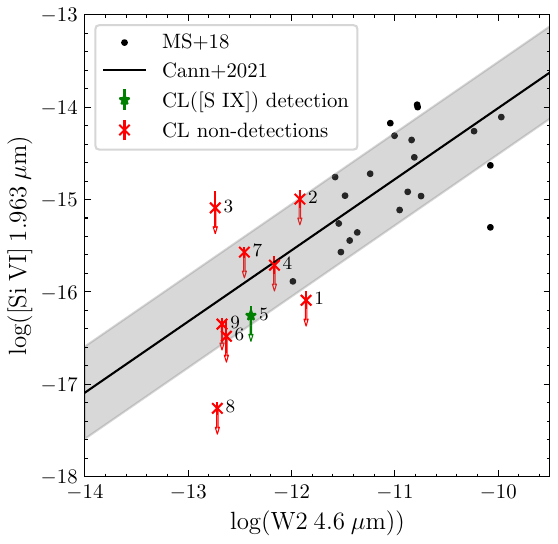}{0.5\textwidth}{}}
\caption{\footnotesize Plot of [\ion{Si}{6}] vs.\ W2 flux for the [\ion{Si}{6}] non-detections in our sample. The solid line indicating the correlation is from \citet{2021ApJ...912L...2C}, which was derived from observations from \citet{2018ApJ...858...48M} and the shaded region indicates the scatter in the data. The targets are marked as J0847 (1), J0928 (2), J1032 (3), J1144 (4), J1205 (5), J1320 (6), J1358 (7), J1428 (8), J1612 (9). Targets whose upper limits are above the relation would require deeper observations in order to detect the [\ion{Si}{6}] line, while for the targets with upper limits below the relation, it is likely that the [\ion{Si}{6}] line is not present. For more details regarding the upper limits, refer to the appendix}
\label{fig:fig4}
\end{figure}

We did not detect any other coronal lines apart from [\ion{S}{9}] in J1205. It is possible that circumnuclear stellar populations dominate the NIR continuum and thus drown out the  CL emission, making fainter lines more difficult to detect \citep{2011ApJ...743..100R}, especially since J1205 has been classified as a compact star-forming galaxy with very young ages ($<$ 3 Myr) for star formation bursts. There is a feature near the noise limit that is suggestive of [\ion{Si}{6}] (see Fig. \ref{fig:fig14}), but the feature did not have sufficient signal to be fit using BADASS. We independently fit a Gaussian to this feature and determine the flux to be 1.78 $\pm$ 0.88 $\times$ 10$^{-17}$ erg cm$^{-2}$ s$^{-1}$, which is below the 3$\sigma$ limit (see Fig. \ref{fig:fig4}). Thus, this feature does not appear to be a 3$\sigma$ detection, but it is possible that with deeper observations, this feature can be detected with greater confidence.

While the detection of the coronal line [\ion{S}{9}] alone and the absence of [\ion{Si}{6}], the most commonly observed coronal line in galaxies \citep{2011ApJ...743..100R,2017MNRAS.467..540L,2018ApJ...858...48M,2021ApJ...911...70B}, may seem surprising, such occurrences are not unprecedented. There have been a few instances of galaxies that have a [\ion{S}{9}] detection, but only upper limits to the [Si VI] flux are reported \citep{2017MNRAS.467..540L, 2022ApJS..261....7D}. Comparing the [\ion{Si}{6}] to [\ion{S}{9}] flux ratios recorded in the literature \citep{2011ApJ...743..100R,2017MNRAS.467..540L,2021ApJ...911...70B,2022ApJS..261....7D}, which range from 0.1 to 8 (more than half the galaxies have ratios less than 3.5), our measured ratio of 0.25 (assuming the 3$\sigma$ flux limit for [\ion{Si}{6}]) falls within this observed range. It is possible that we simply did not obtain sufficiently deep observations to detect the [\ion{Si}{6}] line in J1205.

\begin{figure*}
\gridline{\fig{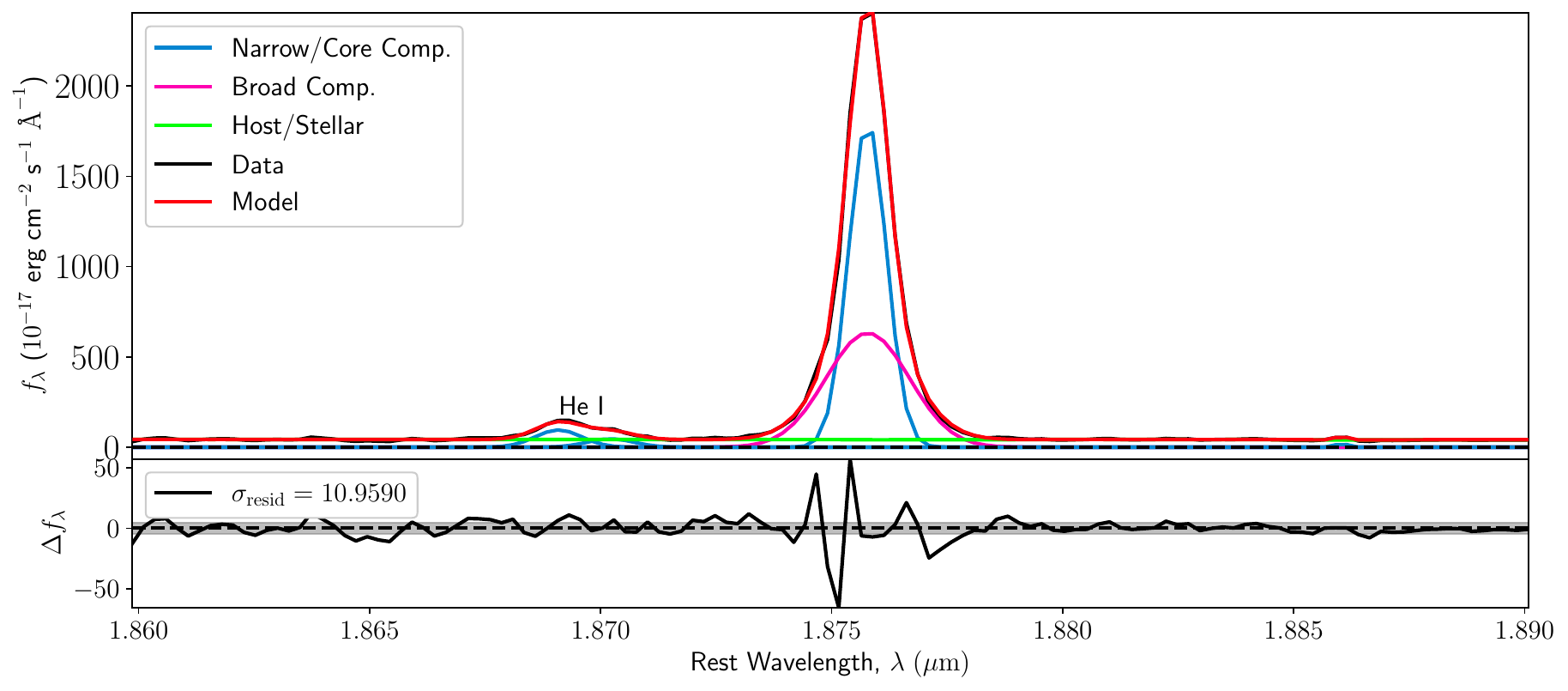}{0.9\textwidth}{(a)}}
\gridline{\fig{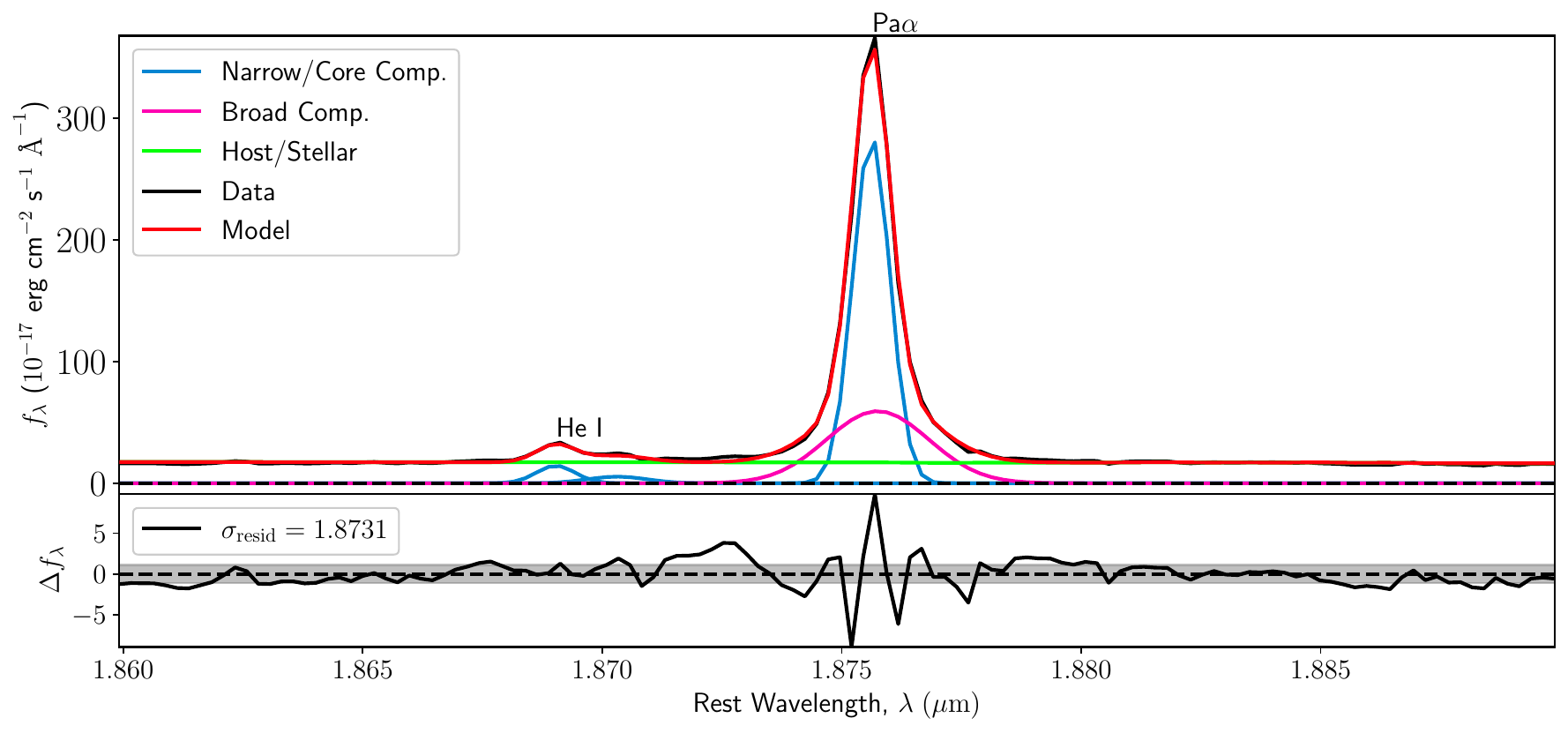}{0.9\textwidth}{(b)}}
\caption{\footnotesize Broad-line fits to the Pa$\alpha$ line in two variable targets, J0928 (a) and J1144 (b), which could indicate a broad component from the BLR of the galaxies. There was an unidentified line next to \ion{He}{1} line at 1.869 $\mu$m, which was fit but left unmarked. The lower panels in both figures depict the residuals to the fit, with the 1$\sigma$ error from the noise in the spectrum shaded in grey around the dashed line indicating 0. \label{fig:fig5}}
\end{figure*}

No coronal lines were detected in the remaining eight targets observed with NIRES, potentially due to the relatively low S/N of the data. To investigate this, we plotted the upper limits to the [\ion{Si}{6}] flux for each target versus the $WISE$ W2 (4.6 $\mu$m) flux in Fig.~\ref{fig:fig4}. We determined the upper limits to the [\ion{Si}{6}] fluxes in the non-detections by estimating the 1$\sigma$ noise levels in the region around [\ion{Si}{6}] flux and multiplying it by three to obtain the 3$\sigma$ upper limit. The solid line in Fig.~\ref{fig:fig4}indicates the best-fit relation, which was derived based on observations from \citet{2018ApJ...858...48M} (indicated as black dots) by \citet{2021ApJ...912L...2C}. The plot indicates that J1428 is the only target for which [\ion{Si}{6}] emission is highly unlikely to be present, as its upper limits are comfortably below the scatter of the relation, and it would have been detected if it were present. For the rest of the targets, we would likely require deeper observations in order to detect [\ion{Si}{6}].  The absence of CLs in these targets is unlikely to be due to high extinction, as J1205 (the target with CL detection) does not have significantly lower extinction (calculated from the H$\alpha$ over H$\beta$ Balmer decrement) than the rest of the observed targets.

While the presence of CLs in a galaxy's spectrum confirms the existence of a black hole, the absence of CLs does not mean the contrary. \citet{2021ApJ...911...70B} analyzed a sample of 9 dwarf galaxies with optical and NIR line ratios indicative of  AGN and found NIR CLs only in 5 of them (55 \%). Several other works in the literature have found less than 100\% CL detections in samples of Sy1 and Sy2 galaxies with similar spectral windows: 66\% (36 out of 54, \citet{2011ApJ...743..100R}), 25\%, (5 out of 20, \citet{2015ApJS..217...13M}), and 43\% (44 out of 102, \citet{2017MNRAS.467..540L}). 

The relatively low detection rates in all of these samples could be an observational effect, as ground-based observations are always hampered by telluric absorption from the atmosphere. Depending on a target's redshift, the CLs can be shifted into regions of strong telluric absorption and/or in regions dominated by strong OH skylines, making it difficult to detect weak CLs. Other factors include a dominant contribution from the stellar population to the NIR continuum and the presence of stellar absorption features.

\vspace*{-2mm} 
\subsection{Detection of broad Pa$\alpha$ lines} \label{sec:sec3.3}

Two of the variable targets (J0928 and J1144) show a broad line component in Pa$\alpha$ at 1.876~$\mu$m. We justify the need for a second component based on the F-test. The F-test determines the significance of the fit between a complex (higher-order) and a simple (lower-order) model in order to justify the number of Gaussian components we include in our fit. To do this, we calculated the standard deviation of the residuals of the fit (given by $\sigma$). We performed the F- test, given as F = ($\sigma_{\mathrm{lower-order}})^2$/($\sigma_{\mathrm{higher-order}})^2$, for one versus two components for the Gaussian fits to the Pa$\alpha$ line. The F-test was calculated over a wavelength range of 1.86–1.89 $\mu$m, which includes the Pa$\alpha$ $\lambda$ 1.876 $\mu$m. An F value greater than 3 indicates that the higher-order fit (i.e., the fit using more components) is justifiable. Using this method, we determined that the Pa$\alpha$ emission lines in J0928 and J1144 both required a second Gaussian component, which we allowed to freely vary to fit the broad feature present in the emission lines using BADASS. The fits to the spectrum can be seen in Fig.~\ref{fig:fig5}.

Both J0928 and J1144 show a broad component to Pa$\alpha$ while not showing a broad component in the optical Balmer lines (at least at the spatial resolution of the SDSS spectrum, observed with a fiber that includes significant contaminating/diluting stellar light). In galaxies with strong dust obscuration, it is possible that the broad line region (BLR) does not manifest itself in optical emission lines. However, NIR wavelengths are less affected by dust extinction, making it easier to observe broad components that would be obscured at optical wavelengths. This method has been used to uncover optically hidden AGN in massive galaxies \citep{1997ApJ...477..631V, 2017MNRAS.467..540L, 2020ApJ...899...82B}. There were no indicators of a broad component being present in any of the forbidden lines in the NIR for the two targets. Combined with the fact that the SDSS spectrum of the two targets also did not have a significant broad component in the forbidden [\ion{O}{3}] lines, it is unlikely that the broad component is produced by an outflow.

\begin{figure}
\gridline{\fig{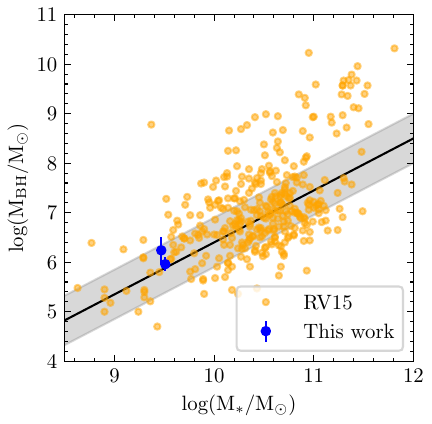}{0.5\textwidth}{}}
\caption{\footnotesize Black hole vs. stellar mass for a sample of galaxies at low-redshift. The yellow points are galaxies at a redshift $<$ 0.06 and include dwarf galaxies with broad lines from \citet{2015ApJ...813...82R} (RV15). The solid line is the expected linear regression from Equations 4 and 5 in RV15, along with the 0.5 dex error, indicated by the gray shaded region. The two blue points indicate the calculated BH masses based on the broad Pa$\alpha$ for J0928 and J1144, respectively, which fits well with the expected black hole mass for the two galaxies. \label{fig:fig12}}
\end{figure}

Assuming the broad lines measured are from the BLR, we used the FWHM and luminosity of the broad Pa$\alpha$ component to estimate black hole masses. Using virial mass estimators from \citet{2018ApJS..238...37K}, we determined the black hole mass (Table \ref{tab:table3}) from the following equation:
\begin{equation}
\frac{M}{M_{\odot}} = 10^{7.07} \left( \frac{L_\mathrm{Pa \alpha}}{10^{42}~\mathrm{erg~s}^{-1}} \right)^{0.49} \times \left( \frac{\text{FWHM}}{10^{3}~\mathrm{km~s}^{-1}} \right)^{2}
\end{equation}

\begin{table}
    \centering
    \caption{Black hole masses}
    \begin{tabular}{lcccccl}
        \hline
        \hline
         & $\mathrm{L_{Pa \alpha}}$  & FWHM &  log$_{10}$($\text{M}_{\text{BH}}$)  & log$_{10}$($\text{M}_{\text{stellar}})$ \\
         & (erg s$^{-1}$) & (km s$^{-1}$)& (M$_{\odot}$)  & (M$_{\odot}$) \\
        \hline
        J0928 & 10$^{42.5 \pm 0.2}$  & 325 $\pm$ 5 & 6.24 $\pm$ 0.27 & 9.47 \\
        J1144 & 10$^{41.2 \pm 0.12}$  & 457 $\pm$ 5 & 5.96 $\pm$ 0.14 & 9.51\\
        \hline
    \end{tabular}
    \label{tab:table3}
\end{table}

The errors given in Table \ref{tab:table3} are the standard deviation of the posterior distribution from the MCMC fitting of the lines. The measured black hole masses agree with the predicted black hole masses based on the stellar masses of J1144 and J0928 (Fig.~\ref{fig:fig12}) from \citet{2015ApJ...813...82R}.

\begin{figure*} [ht!]
\gridline{\fig{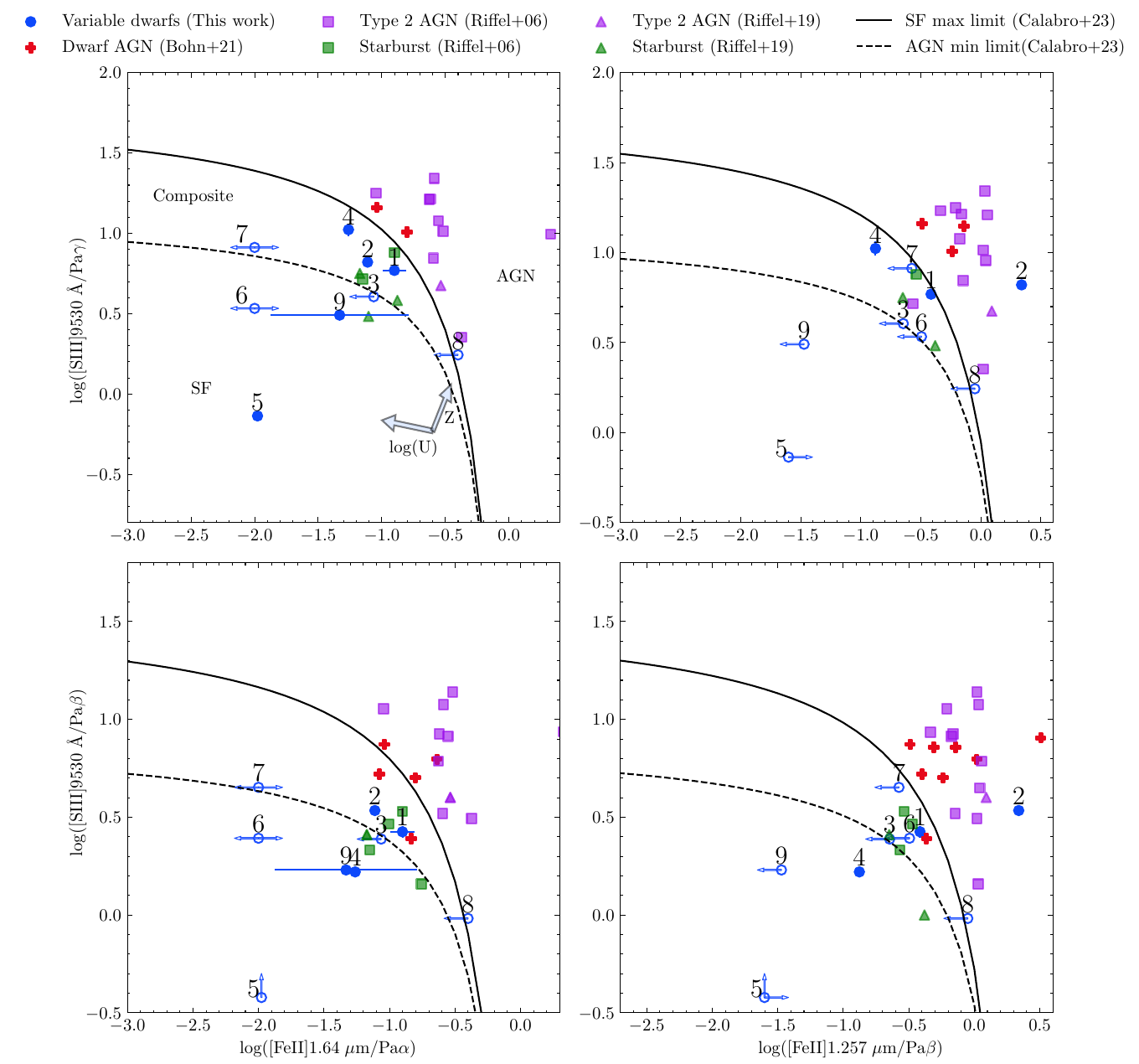}{1\textwidth}{}}
\caption{\footnotesize NIR diagnostic diagrams using AGN and SF limits from \citet{2023A&A...679A..80C}. The targets are marked as J0847 (1), J0928 (2), J1032 (3), J1144 (4), J1205 (5), J1320 (6), J1358 (7), J1428 (8), J1612 (9). For the targets where some of the line fluxes were not available, we plotted them as open circles with arrows, indicating upper and lower limits. Dwarf Galaxies with AGN from \citet{2021ApJ...911...70B} and Type 2 AGN from \citet{2006A&A...457...61R} are plotted in red and green respectively. The gas-phase metallicity (Z) increases from bottom to top, and the ionization parameter (log (U)) increases from right to left. Four of the targets moved to the composite/AGN region across all four NIR diagnostic diagrams, as compared to their positions in the SF region in the optical diagnostic diagrams.
\label{fig:fig6}}
\end{figure*}

\subsection{NIR diagnostic diagrams} \label{sec:sec3.4}

Optical diagnostic diagrams like the BPT diagram have been well studied and used extensively to distinguish between AGN and SF regions in galaxies. However, similar diagnostic diagrams in the NIR are scarce, chiefly due to a lack of large spectroscopic surveys in the infrared. Nevertheless, there have been several attempts to recreate similar diagnostic diagrams in the infrared that can be used to distinguish between AGN and SF regions \citep{1998ApJS..114...59L, 2014ApJ...794..112S, 2015A&A...578A..48C}. In particular, \citet{2023A&A...679A..80C} used CLOUDY photoionization models to identify AGN - SF diagnostics based on the ratio of bright near-infrared emission lines such as  [\ion{S}{3}] 9530 \AA, [\ion{C}{1}] 9850 \AA, [\ion{P}{2}] 11880 \AA, [\ion{Fe}{2}] 12570 \AA, and [\ion{Fe}{2}] 16470 \AA\ to the Paschen lines. They used a sample of 130 AGN and SF galaxies with redshifts $0<z<3$ observed using JWST-NIRSpec \citep{2022A&A...661A..80J} as part of the Cosmic Evolution Early Release Science Survey \citep[CEERS; PID 1345, PI Finkelstein,][]{2023ApJ...946L..13F} and found that this diagram is able to successfully distinguish between AGN and SF galaxies. Furthermore, they found 60\% more AGN using NIR diagnostic diagrams than when using optical BPT diagrams. Thus NIR diagnostic diagrams are able to uncover optically `hidden' AGNs, thus highlighting the importance of using NIR wavelengths to potentially locate AGN in dusty galaxies. Since optical emission lines are significantly affected by extinction, it is possible to overlook AGN indicators from optical emission lines due to extinction. On the other hand, NIR emission lines are less sensitive to dust and thus could prove to be more helpful in identifying AGN activity in dusty galaxies.

Most of the MIR variable targets in our sample fall in the SF region of the optical BPT diagram as seen in Fig.~\ref{fig:fig2}. We plotted them in the various NIR diagnostic diagrams in Fig.~\ref{fig:fig6}. We primarily used the ratios of [\ion{S}{3}] $\lambda$9530 \AA/Pa$\gamma$ and [\ion{S}{3}] $\lambda$9530 \AA/Pa$\beta$ to [\ion{Fe}{2}]  $\lambda$16470 \AA/Pa$\alpha$ and [\ion{Fe}{2}] $\lambda$12570 \AA/Pa$\beta$, obtained from NIRES. For four of the targets (marked as filled circles in Fig.~\ref{fig:fig6}), these were the lines with signal-to-noise ratios (S/N) $>$ 3 in our sample, and their flux values could be estimated with minimum uncertainties. For the remaining targets, we could only determine one of the ratios as at least one of the emission lines on the x-axis ([\ion{Fe}{2}] 1.64$\mu$m, [\ion{Fe}{2}] 1.25$\mu$m, Pa$\alpha$ or Pa$\beta$) fell in the telluric region or were in a noisy part of the spectrum, and thus we could not measure the fluxes. We instead find the upper limits (if the [\ion{Fe}{2}] lines were undetermined) or lower limits (if the Paschen lines were undetermined) to the emission line ratios and showed the targets as open circles with arrows to indicate upper or lower limits (both if neither of the line fluxes used in the ratio could be determined) to the ratios in Fig.~\ref{fig:fig6}. The upper limits for J1032, J1428, and J1320 did not exclude the possibility that these targets are AGN.

We found that four of the variable targets (J0847, J0928, J1144, and J1358) fell in the composite and AGN regions across all four NIR diagnostic diagrams. J0928, which has a broad Pa$\alpha$ line, moved from the SF region in the optical to the AGN region of two of the NIR diagnostic diagrams. The other broad line target, J1144, also moved into the composite region of all four NIR diagnostic diagrams. 

While NIR diagnostic diagrams appear to be more effective in some cases at identifying obscured AGN in dwarf galaxies, they are still limited by the gas-phase metallicities of the galaxies. J1205, with a coronal line indicating the presence of a black hole, still lies firmly in the SF region of all the NIR diagnostic diagrams.

\subsection{Ruling out supernova contribution to variability} \label{sec:sec5.4}
Although the nine galaxies that were observed using NIRES are determined to exhibit AGN-like variability from the light curves, our selection method does not completely rule out contributions from SN. Five of the galaxies have alternate evidence of AGN activity based on the presence of coronal lines, broad Pa$\alpha$ lines, and emission line ratios indicative of AGN, which makes it unlikely that there is a significant contribution from SN to the variability.  However, \citet{2016ApJ...829...57B} showed that in the optical, broad H$\alpha$ lines could be caused by SN activity. The ideal method to rule out the possibility of broad lines being caused by SN is to obtain multi-epoch spectroscopy that would show whether the broad emission lines persist. Although we do not have multi-epoch spectra and cannot check for line variability, we did inspect all the light curves by eye, and determined that they do not show the rapid increase and decline in their curves that is characteristic of supernovae. Furthermore, we inspected the NIR spectrum for the common NIR emission lines associated with SN \citep{2018ApJ...864L..20R} and did not find any features that indicated the presence of SN.

\subsection{Summary of results}

\begin{figure*}
\gridline{\fig{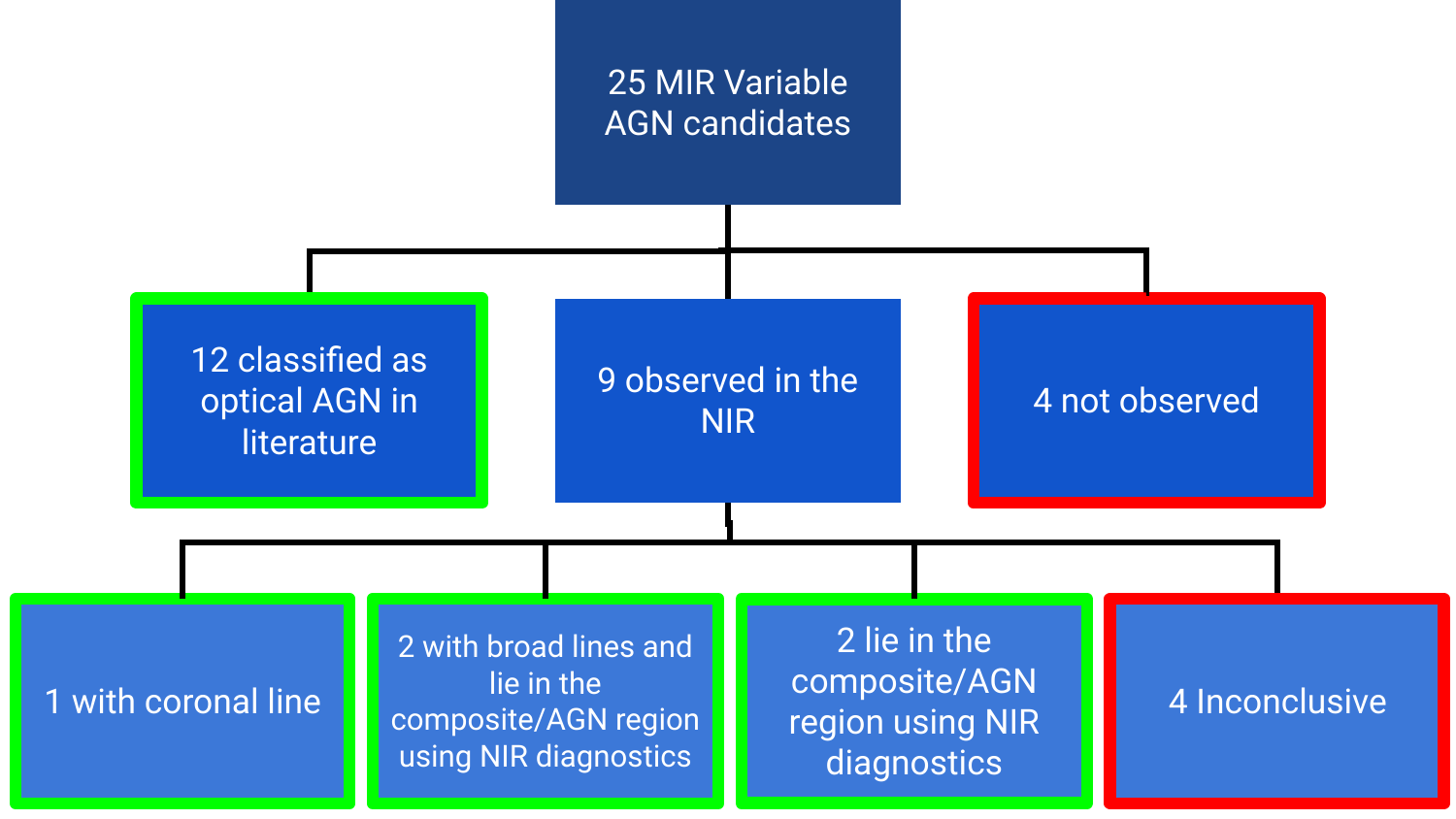}{0.8\textwidth}{}}
\caption{\footnotesize Flowchart summarizing the results characterizing the effectiveness of MIR variability in tracing AGN activity in low mass galaxies. \label{fig:fig13}}
\end{figure*}

Dwarf galaxies with active black holes have been found to not display AGN signatures uniformly across different detection methods. If we have a dwarf galaxy that is variable in the MIR, how reliably can we determine that it is actually hosting an active black hole? In our work, we found that out of 25 dwarf galaxies that exhibit MIR variability, we have independent evidence that 17 of them (68\%) host an AGN, strongly suggesting that MIR variability is due to the presence of an AGN in these targets. Twelve of these were classified as optical AGN, and the remaining five had NIR signatures of hosting an AGN. Four did not have any optical indicators of AGN, but we were not able to obtain NIR observations for them, so the true fraction of independently confirmed AGN in the sample of MIR variable AGN can be higher than 68\% (see Fig.~\ref{fig:fig13} for a summary). 

Our results showed that if a galaxy is variable in the MIR, there is a high probability that it is actually hosting an active black hole. Using TESS light curves, \citet{2023MNRAS.525.5795T} identified a similar fraction (62\%) of optically variable galaxies that were either previously classified as AGNs in the literature or confirmed as AGNs based on a combination of emission-line diagnostics, mid-IR colors, or X-ray luminosity in their analysis. Thus, variability is a good indicator of AGN activity in dwarf galaxies and can be used to recover a larger fraction of AGN that is missed by other methods.

Signs of active black holes in dwarf galaxies are often difficult to distinguish from those originating from SF activity. The low luminosity of the AGN can lead to the possibility that SF activity capable of producing similar luminosities often act as interlopers. In the infrared in particular, \citet{2016ApJ...832..119H} found that most of the BPT-confirmed AGNs from \citet{2013ApJ...775..116R} were not selected as $WISE$ AGN, which could be due to the host galaxy light dominating in the mid-infrared color space. Conversely, star-forming dwarf galaxies are capable of heating dust in such a way that mimics the infrared colors of more luminous AGNs. That is why $W1-W2$ colors alone should not be used to select for AGN in dwarf galaxies.

\begin{figure*}
\gridline{\fig{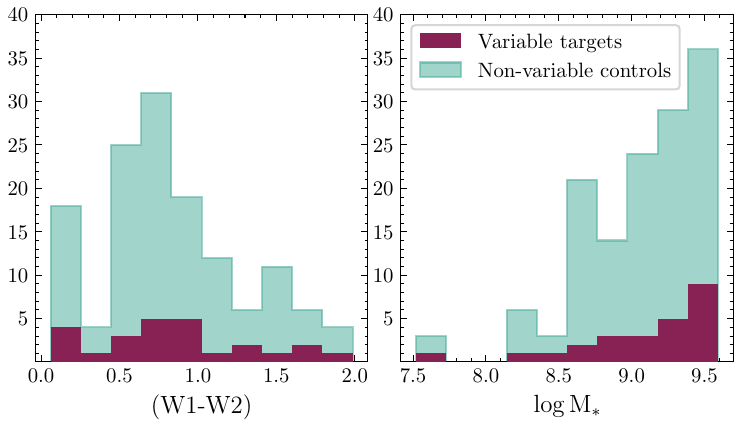}{1\textwidth}{}}
\caption{\footnotesize Histograms showing the distribution of the $WISE$ $W1-W2$ colors and stellar masses of the target sample of variable dwarf galaxies and the control sample of non-variable dwarf galaxies. \label{fig:fig7}}
\end{figure*}
\section{Comparison with non-variable dwarf galaxies}
\label{sec:sec4}

Given the potential contamination from SF processes at infrared wavelengths, it is crucial to test whether MIR variability is accurately tracing AGN activity or if it is being caused by SF interlopers in dwarf galaxies. We have found that MIR variability is frequently accompanied by alternate indicators of AGN (see Sec. \ref{sec:sec2.1} and Sec. \ref{sec:sec4}), making it less likely that it is caused by stellar processes. Additionally, we compared dwarf galaxies that exhibit MIR variability with similar dwarf galaxies that do not exhibit MIR variability. We did this to recover differences in optical and infrared galaxy properties, if any, between the two samples of galaxies and understand the effects of any potential biases caused by host galaxy properties on the observed MIR variability.

\subsection{Control sample selection}
We selected a control sample of non-variable dwarf galaxies, which were matched in $WISE$ $W1-W2$ colors and stellar mass with the sample of variable dwarf galaxies. We obtained five control galaxies on average for each of the 25 target galaxies that we initially classified as variable and thus created a control sample of 136 non-variable dwarf galaxies. A two-sample Kolmogorov-Smirnov (K-S) test between the $W1-W2$ colors and stellar masses of the variable galaxies and non-variable control galaxies gives p=0.99 and p=0.56, respectively, indicating that the control galaxies do not have systematically different $WISE$ colors or stellar masses (see Fig.~\ref{fig:fig7}). None of the galaxies in the control sample passed the variability tests based on our initial constraints (see Section 2.1), and thus can be safely assumed to be non-variable. The two samples have similar redshift distributions, ranging from 0.01 to 0.1.

\begin{figure*}
\gridline{\fig{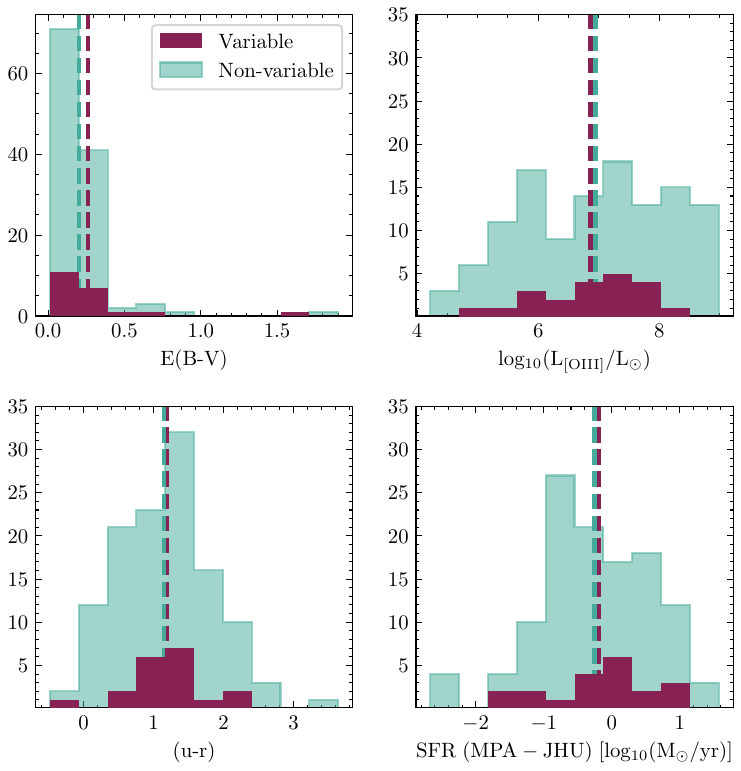}{1\textwidth}{}}
\caption{\footnotesize Histograms showing the comparison optical properties of the variable and the non-variable dwarf galaxies. The dashed lines indicate the average values for the two samples. We did not find a significant difference in several key optical properties, such as extinction, [\ion{O}{3}] luminosity, optical colors and SFRs.
\label{fig:fig8}}
\end{figure*}

We then obtained emission line flux values for [\ion{O}{2}] $\lambda \lambda$3727,3729 \AA, \ion{He}{2} $\lambda$4686, H$\beta$ $\lambda$4861 \AA, [\ion{O}{3}] $\lambda$ 5007 \AA, [\ion{O}{1}] $\lambda$ 6300 \AA, [\ion{N}{2}] $\lambda$6584 \AA,  H$\alpha$ $\lambda$6563 \AA\ and [\ion{S}{2}] $\lambda$6731 \AA\ from the MPA-JHU catalog \citep{2003MNRAS.341...33K}. We measured extinction values for individual galaxies from the H$\alpha$/H$\beta$ Balmer decrement. We then obtained extinction corrected [\ion{O}{3}] $\lambda$5007 \AA\ luminosity values for all the galaxies in the target and the control samples. We also measured the star-formation rates (SFRs) for the galaxies using a number of methods. We directly obtained the SFRs for the galaxies from the MPA-JHU catalog, which were calculated based on the methods used in \citet{2004MNRAS.351.1151B}. We independently estimated SFRs from the extinction-corrected H$\alpha$ luminosity values based on the relations by \citet{1998ARA&A..36..189K} as well as from the extinction-corrected [\ion{O}{2}] fluxes using Equation 11 in \citet{2004AJ....127.2002K}. The SFRs calculated by these different optical methods agreed with each other. We determined the optical colors (u-r) from SDSS for the different galaxies. 

\begin{figure}
\gridline{\fig{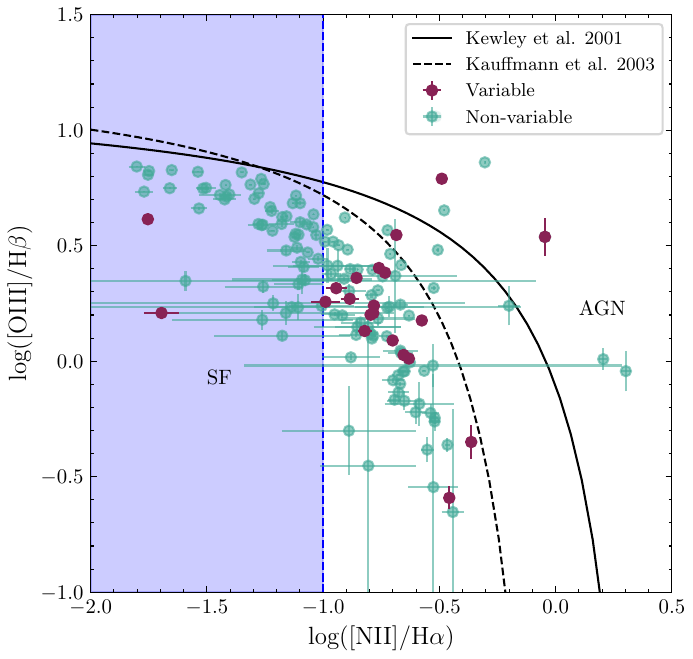}{0.5\textwidth}{}}
\caption{\footnotesize The BPT diagram of the variable and non-variable samples. Galaxies with log [\ion{N}{2}]/H$\alpha$ $<$ -1 (indicative of low metallicity) are in the blue shaded region). Using log [\ion{N}{2}]/H$\alpha$ as a proxy for metallicity, we found that a larger fraction of variable dwarfs have higher metallicities than non-variable dwarfs.
\label{fig:fig10}}
\end{figure}

\subsection{Large-scale properties of variable and non-variable dwarf galaxies} \label{sec:sec4.2}

Comparing various properties of variable and non-variable targets with similar $W1-W2$ colors, we did not find a significant difference between the extinction values, [\ion{O}{3}] luminosities, optical colors, and SFRs between both samples (Fig.~\ref{fig:fig8}). 

Using the emission line values obtained from the MPA-JHU catalog, we plotted the BPT diagrams of both samples (see Fig.~\ref{fig:fig10}). We found that 3/25 (12\%) of the variable targets lie above the \citet{2003MNRAS.346.1055K} demarcation line for the AGN region, while 8/136 (0.05\%) of the non-variable controls lie in the AGN region.

We obtained \ion{He}{2} $\lambda$4686 emission line fluxes for eight of the variable galaxies and eighty-two of the non-variable galaxies from the MPA-JHU catalog. No significant difference was found between the positions of the variable and non-variable galaxies in the \ion{He}{2}$\lambda$4686/H$\beta$ versus [\ion{N}{2}]$\lambda$6584/H$\alpha$ diagnostic diagram presented in \citet{2012MNRAS.421.1043S}. While some of the variable galaxies were previously identified as AGNs using this diagnostic, we identified four additional variable galaxies that meet the AGN criteria. These galaxies are marked with blue checkmarks in Column 5 in Table \ref{table:Table1}.

Determining gas-phase metallicities for the targets is challenging as the traditional methods that depend on various emission lines such as [\ion{O}{3}] and [\ion{O}{2}] often do not account for the contamination due to AGN activity. Moreover, studies have shown that there is an ambiguity in determining metallicities using methods such as the ratio of [\ion{O}{2}]$\lambda\lambda$ 3726,3729 + [\ion{O}{3}]$\lambda\lambda$4959,5007 over H$\beta$  \citep[R23,][]{1979MNRAS.189...95P}, particularly in the range of metallicities (7.5 $<$ 12 + log(O/H) $<$ 8.5) that we would expect these dwarf galaxies to have \citep{2008A&A...488..463M, 2015ApJ...813..126J}. Thus, to comment on the metallicities of the targets, we looked solely at the [\ion{N}{2}]/H$\alpha$ line ratios from the BPT diagram, which can be used as a proxy for metallicity \citep{2002ApJS..142...35K, 2004MNRAS.348L..59P, 2008ApJ...681.1183K}. 

We found that, on average, variable targets have higher metallicities (log [\ion{N}{2}]/H$\alpha$ $>$ -1) as compared to the non-variable targets. Only 2/25 (8\%) targets have [\ion{N}{2}]/H$\alpha$ lower than -1 as compared to 59/136 (43\%) for the control sample, as seen in Fig.~\ref{fig:fig10}.  BPT-selected AGNs in dwarf galaxies are expected to be biased towards higher metallicities \citep{2013ApJ...775..116R}. However, we adopted a completely independent method of selecting AGN in dwarf galaxies, and we still found that variable dwarf galaxies (tracing the AGN) have higher metallicities than non-variable dwarfs. Since AGN contribution can also lead to an increase in the [\ion{N}{2}]/H$\alpha$ ratio, a large fraction of variable targets with high [\ion{N}{2}]/H$\alpha$ ratios may be a direct consequence of the presence of AGNs in the variable targets.

\subsection{Optical broad lines and $W2-W3$ colors}
\begin{figure}
\gridline{\fig{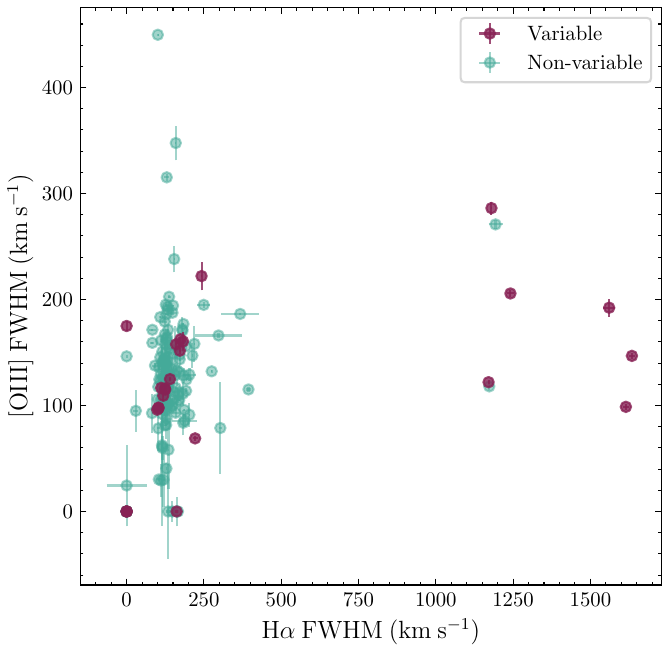}{0.5\textwidth}{}}
\caption{\footnotesize Plot of the H$\alpha$ FWHM vs the [O\,III] FWHM for the variable and non-variable targets. A larger fraction of variable targets have very broad (FWHM $>$ 500 km s$^{-1}$) Balmer lines than non-variable targets.\label{fig:fig11}}
\end{figure}

We obtained the SDSS spectra of all the galaxies in the target and the control samples. We then performed spectral fits for the entire sample in the regions that include the prominent Balmer lines (4400 - 7000 \AA) and fit simultaneously for  H$\beta$ $\lambda$ 4861 \AA, [\ion{O}{3}] $\lambda\lambda$ 4963, 5007 \AA, H$\alpha$ $\lambda$ 4861 \AA, and [\ion{N}{2}]  $\lambda\lambda$ 6548, 6583 \AA. We found that variable targets have broader Balmer lines compared to non-variable targets (Fig.~\ref{fig:fig11}): 
6/25 (24\%) targets have H$\alpha$ FWHM $>$ 500 km s$^{-1}$ while only 2/136 (0.015\%) of the controls have H$\alpha$ FWHM $>$ 500 km s$^{-1}$. Broad Balmer lines could indicate the presence of AGN activity, but there could be potential contamination from star formation. We did not find any broad outflowing component in the forbidden [\ion{O}{3}] lines of both samples, indicating that neither sample has strong outflows in ionized gas.

This higher incidence of optical broad lines in the sample of galaxies exhibiting MIR variability as compared to non-variable galaxies is surprising. The presence of broad lines in the spectrum of AGN is typically attributed to an orientation effect. In type 1 AGNs, the nucleus is oriented so that we can observe the BLR and narrow line regions (NLR). MIR variability, on the other hand, should have no dependence on orientation, as the variability is believed to originate from the dust surrounding the accretion disk and is expected to be isotropic. 
However, \citet{2023ApJ...958..135S} found that low-luminosity type 2 AGNs tend to have lower variability amplitudes than their type 1 counterparts. They attribute this to the type 2 AGN having distinctive central structures due to their low luminosity or their MIR brightness being contaminated by emission from the cold dust in the host galaxy. On the other hand, recent observations of extended (tens to hundreds of parsecs) MIR emission from the polar regions of AGNs suggest that MIR emission may be dominated by dust from the polar regions rather than from the torus \citep{2016ApJ...822..109A, 2019MNRAS.489.2177A}. This interpretation possibly suggests that we can simultaneously observe the BLR as well as MIR emission. Our observations of a large number of Type 1 AGN that are variable in MIR can be explained with either of these scenarios.

\begin{figure}
\gridline{\fig{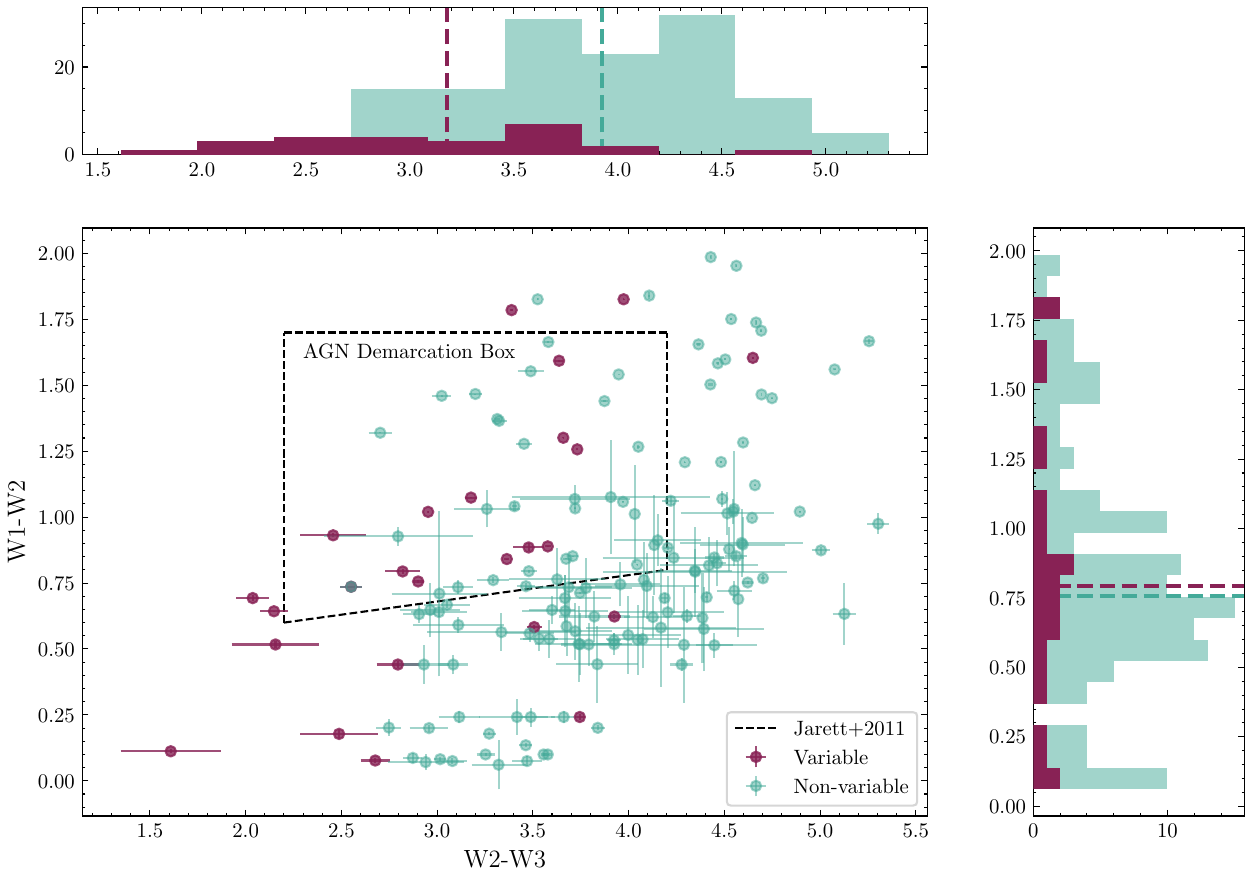}{0.5\textwidth}{}}
\caption{\footnotesize The $WISE$ color-color diagram for the variable and non-variable galaxies, with histograms plotting the distribution of the values in $W1-W2$ (right) and $W2-W3$ (top). Although both samples have similar $W1-W2$ colors, non-variable galaxies have redder $W2-W3$ colors.
\label{fig:fig9}}
\end{figure}

We also obtained $WISE$ $W1$, $W2$, $W3$, and $W4$ fluxes for all the galaxies. Examining the infrared colors of both samples (Fig.~\ref{fig:fig9}), we found that non-variable targets have higher values of $W2-W3$ (redder $W2-W3$ colors) on average than variable targets (3.92 vs 3.18). Redder $W2-W3$ colors indicate activity caused by starburst events \citep{2013AJ....145....6J}. A p-value of 0.027 ($<$ 0.05) indicates that there is a statistically significant difference in $W2-W3$ between the variable and control samples. This suggests that variable targets have bluer $W2-W3$ colors, indicating that there is likely lesser contribution from dust heated by star formation in the MIR for the variable targets. This difference in the contribution of SF activity is not apparent from the optical, where both samples have similar SFRs (see Fig. \ref{fig:fig8}). Even with similar $W1-W2$ colors, variable dwarfs still have bluer $W2-W3$ colors. Thus, it is likely that AGN activity causes redder $W1-W2$ colors in variable targets, while for the non-variable galaxies, the contribution stems from strong star formation activity.

\section{Conclusions} \label{sec:sec5}
In this work, we selected a sample of twenty-five dwarf galaxies that are identified to be variable in the MIR from the light curves from the most recent release of the All$WISE$/NEO$WISE$ catalog. We determined that 12/25 (48\%) of these galaxies had been confirmed as AGN in the literature by several optical diagnostic methods, such as the BPT diagram, presence of broad lines or strong He\,II emission, indicating that MIR variability accurately traces black hole activity in these small dwarf galaxies. For a subsample of the variable dwarf galaxies that did not have any optical confirmations of AGN activity, we obtained NIR observations using Keck/NIRES. Our main results can be summarized as follows (also see Fig.~\ref{fig:fig13}):

\begin{enumerate}
    \item We present the first detection of a NIR coronal line ([S IX]) in J1205, which has the lowest mass (log(M$_{*}$) = 7.5 M$_{\odot}$) among all the objects in our sample. Most commonly used AGN diagnostics, such as BPT diagrams and He II emission lines, fail to detect the presence of an AGN in J1205, thus highlighting that  MIR variability can be a powerful and sometimes the only tool to detect AGNs in low-mass galaxies.
    \item We find evidence of a broad Pa$\alpha$ line potentially from the BLR in two of the variable targets (J0928 and J1144). These targets lie in the SF region of the optical BPT diagram, but they lie in the composite or AGN regions of the NIR diagnostic diagrams. The black hole masses implied by their broad lines are consistent with the black hole to stellar mass relation from RV15.
    \item Using NIR diagnostic diagrams, which make use of NIR emission line ratios in order to differentiate between AGN and SF regions in galaxies, we find that several of the variable targets (4/9) that we observed moved into the AGN/composite region of the diagnostic diagram compared to their location in the SF regions of the optical BPT diagram. 
    \item Comparing a sample of variable and non-variable dwarf galaxies with similar stellar mass and $WISE$ $W1-W2$ colors, we find that variable dwarfs have bluer $W2-W3$ colors than non-variable dwarfs. The redder $W1-W2$ colors of non-variable dwarfs are likely because of high star formation. Thus, variable dwarfs accurately trace AGN activity independent of mid-infrared colors.
    \item Variable dwarfs have a higher fraction of galaxies that have broad ($>$ 500 km s$^{-1}$) Balmer (H$\beta$ and H$\alpha$) lines than the non-variable dwarfs (24\% compared to 0.015\%).  While this could simply indicate a higher incidence of AGN, it is remarkable that such a large fraction would be type 1 AGN. 
\end{enumerate}
We find independent confirmation of AGN activity in at least 68 \% out of 25 MIR variable dwarf galaxies, indicating that MIR variability can be used to accurately detect AGN in dwarf galaxies. Improving methods to detect low-mass active black holes is the first step in constraining the black hole occupation fraction in dwarf galaxies, thus determining the evolutionary pathways for black hole seeds in the early universe. Our study shows that using indicators in the infrared is crucial to obtaining a more complete census of black holes in dwarf galaxies.  With the increasing amount of infrared data being available from JWST and the upcoming infrared transient survey telescopes such as the Palomar Gattini-IR \citep{2019NatAs...3..109M}, harnessing the power of MIR variability will likely be fruitful in uncovering large numbers of active black holes in dwarf galaxies.

\begin{acknowledgments}
We thank the anonymous referee for their thoughtful feedback and constructive comments that greatly helped to improve this paper. Partial support for this project was provided by the National Science Foundation under grant No.\ AST 1817233. The data presented herein were obtained at the W. M. Keck Observatory, which is operated as a scientific partnership among the California Institute of Technology, the University of California, and the National Aeronautics and Space Administration. The Observatory was made possible by the generous financial support of the W. M. Keck Foundation. The authors wish to recognize and acknowledge the very significant cultural role and reverence that the summit of Maunakea has always had within the indigenous Hawaiian community. We are most fortunate to have the opportunity to conduct observations from this mountain. 

This publication makes use of data products from the Wide-field Infrared Survey Explorer, which is a joint project of the University of California, Los Angeles, and the Jet Propulsion Laboratory/California Institute of Technology, and NEO$WISE$, which is a project of the Jet Propulsion Laboratory/ California Institute of Technology. $WISE$ and NEO$WISE$ are funded by the National Aeronautics and Space Administration.

SDSS-IV is managed by the Astrophysical Research Consortium for the Participating Institutions of the SDSS Collaboration. Funding for the Sloan Digital Sky Survey IV has been provided by the Alfred P. Sloan Foundation, the U.S. Department of Energy Office of Science, and the Participating Institutions. SDSS-IV acknowledges support and resources from the Center for High-Performance Computing at the University of Utah. 
\end{acknowledgments}

\facilities{$WISE$, NEO$WISE$, KCWI/Keck, SDSS}

\software{astropy \citep{2013A&A...558A..33A,2018AJ....156..123A},
          BADASS \citep{2021MNRAS.500.2871S},
          pPXF, \citep{2004PASP..116..138C},
          \textsc{REDSPEC} (\url{https://www2.keck.hawaii.edu/inst/nirspec/redspec.html}),
          SciencePlots \citep{SciencePlots}}

\appendix \label{appendix}
\vspace{-10mm}
\begin{table}
    \centering
    \begin{tabular}{lccl}
        \hline
        \hline
        Target & [\ion{Si}{6}] flux upper limits\\
         & (10$^{-17}$ erg cm$^{-2}$ s$^{-1}$)\\
         \hline
        J0847 & 8.12\\ 
        J0928 & 100.97\\
        J1032 & 81.28\\ 
        J1144 & 19.47\\ 
        J1205 & 5.55\\ 
        J1320 & 3.31\\ 
        J1358 & 27.54\\ 
        J1428 & 0.55\\ 
        J1612 & 4.57 \\
        \hline
    \end{tabular}
    \caption{3$\sigma$ Upper limits to the [\ion{Si}{6}] flux measurements for the non-detections of coronal lines in this galaxies observed with NIRES (See Figure \ref{fig:fig4})}
    \label{tab:table4}
\end{table}

The non-uniform detection of coronal lines across galaxies is likely due to insufficient observation times in order to detect the typically faint coronal lines. We present the upper limits for the [\ion{Si}{6}] flux for the nine galaxies observed with NIRES in this paper (Table \ref{tab:table4}). The upper limits to the [\ion{Si}{6}] fluxes were determined by estimating the 1$\sigma$ noise levels in the region around [\ion{Si}{6}] flux and multiplying it by three to obtain the 3$\sigma$ upper limit indicated in Table \ref{tab:table4}. None of the observed galaxies had [\ion{Si}{6}] detections.




\bibliography{references}{}
\bibliographystyle{aasjournal}



\end{document}